\documentclass[aps,prd,preprint,groupedaddress,showpacs]{revtex4}
\usepackage{epsfig}
\usepackage{graphics}
\usepackage{amssymb,amsmath}
\usepackage{color}

\newcommand{\tr}{\mathop{\mbox{tr}}\nolimits}

\begin{document}

\leftmargin -2cm
\def\choosen{\atopwithdelims..}


 \boldmath

\title{Diphoton production at Tevatron and the LHC in the NLO$^\star$ approximation of the Parton Reggeization Approach.} \unboldmath

\author{\firstname{M.A. }\surname{Nefedov}}
\email{nefedovma@gmail.com} \affiliation{Samara State University,
Ac.\ Pavlov\ st., 1, 443011 Samara, Russia}

\author{\firstname{V.A.} \surname{Saleev}} \email{saleev@samsu.ru}
\affiliation{Samara State University, Ac.\ Pavlov\ st., 1, 443011 Samara,
Russia} \affiliation{S.P.~Korolyov Samara State Aerospace
University, Moscow Highway, 34, 443086, Samara, Russia}

\begin{abstract}
  The hadroproduction of prompt isolated photon pairs at high energies is studied in the NLO$^\star$ framework of the
  Parton Reggeization Approach. The real part of the NLO corrections is computed, and the procedure for the subtraction
   of double counting between real parton emissions in the hard-scattering matrix element and unintegrated PDF is
   constructed for the amplitudes with Reggeized quarks in the initial state. The matrix element of the important NNLO subprocess
    $RR\to \gamma\gamma$ with full dependence on the transverse momenta of the initial-state Reggeized gluons is obtained.
     We compare obtained numerical results with diphoton spectra measured at Tevatron and the LHC, and find
      a good agreement of our predictions with experimental data at the high values of diphoton transverse
      momentum, $p_T$, and especially at the $p_T$ larger than the diphoton invariant mass, $M$.  In this multi-Regge kinematics region, the NLO correction is strongly suppressed, demonstrating the self consistency of the Parton Reggeization
      Approach.
\end{abstract}

\maketitle

\section{Introduction}
\label{sec:intro}
  Nowadays, the inclusive hadroproduction of pairs of isolated prompt photons (diphotons) is a subject of intense experimental and theoretical studies.
   From the experimental point of view, this process forms an irreducible background in the searches of heavy neutral resonances in the
    diphoton decay channel, such as Standard Model Higgs boson~\cite{SM_Higgs_disc} and it's Beyond Standard Model counterparts~\cite{BSM_anal}.
     As for the process
     itself, it allows us to define the set of inclusive differential cross sections over such variables as the invariant mass of the pair
      ($M$), it's transverse momentum ($p_T$), azimuthal angle between transverse momenta of the photons ($\Delta\phi$),
      rapidity of the photon pair ($Y_{\gamma\gamma}$), Collins-Soper angle in the center of mass frame of the photon pair ($\theta$) and a few
      others~\cite{CDF_data_2012}. Most of these spectra are measured with high precision both at  Tevatron~\cite{CDF_data_2012}
      and the LHC~\cite{ATLAS_data_2013}.

  On the theoretical side, providing the predictions for the above mentioned rich set of differential spectra is a challenging
   task even for the state of the art techniques in perturbative Quantum Chromodynamics (pQCD). While, for the inclusive isolated prompt photon production,
   the $p_T$-spectra from CDF~\cite{CDF_1gamma}, ATLAS~\cite{ATLAS_1gamma} and CMS~\cite{CMS_1gamma} are described within experimental
    uncertainties in the Next to Leading Order (NLO) of conventional Collinear Parton Model (CPM) of the QCD~\cite{Jetphox}. Also, the notably
     good results where obtained for these spectra already in the Leading Order (LO) of $k_T$-factorization in the Ref.~\cite{SVAphoton, KSSY_photon_jet}.
     In contrast, existing NLO CPM calculations, implemented in the \texttt{DIPHOX}~\cite{Diphox} Monte-Carlo event generator,
     provide very poor description of $p_T$ and $\Delta\phi$ distributions measured by ATLAS~\cite{ATLAS_data_2013}. In the CPM, the full
     NNLO accuracy is required to provide qualitatively reasonable description of all distributions~\cite{2gNNLO}.

  Part of these difficulties can be traced back to the shortcomings of the CPM approximation, where the transverse momentum of
  initial state partons is integrated over in the Parton Distribution Functions (PDFs), but neglected in the hard scattering part
  of the process. Such treatment is justified for the fully inclusive single scale observables, such as deep inelastic scattering
  structure functions
  or $p_T$-spectra of single prompt photons and jets, where the corrections breaking the collinear factorization are shown to be
  suppressed by a powers of the hard scale~\cite{Collins_QCD}.

  For the multi-scale differential observables, there is no obvious reason why the fixed-order calculation in the CPM should be a good approximation. Usually the simple picture of
  factorization of the cross section of the hard process into the convolution of hard-scattering coefficient and some PDF-like objects is kept,
  but kinematical approximations are relaxed. In the treatment of Initial State Radiation (ISR) corrections
  in the
  Soft Collinear Effective Theory (SCET)~\cite{SCET} or in the Transverse Momentum Dependent (TMD) factorization
  formalism~\cite{TMD_Noevol, CSS_resumm, Collins_QCD},
  the transverse-momentum of the initial-state parton is kept unintegrated on the kinematical level, but neglected in the hard-scattering part,
  which is justified e.~g. when the $p_T$ of the exclusive final state is much smaller than it's invariant mass, so that the following hierarchy
  of the light-cone momentum components for the initial-state parton is preserved: $q^\mp\ll |{\bf q}_T|\ll q^\pm=x\sqrt{S}$.

  In the opposite limit, when $q^\mp\ll |{\bf q}_T|\sim q^\pm=x\sqrt{S}$, the $k_T$-factorization~\cite{kTf} is valid, and transverse momentum
  of the initial state parton can no longer be neglected in the hard scattering amplitudes. To obtain the suitable hard scattering matrix element
  we will apply the hypothesis of parton Reggeization, which will be described below. In what follows we will refer to the combination of
  $k_T$-factorization with hard scattering matrix elements with Reggeized partons in the initial state as the Parton Reggeization Approach (PRA).
  This approach is mostly suitable for the study of the production of the final states with high $p_T$ and small invariant mass in the central
  rapidity region. At high energies $\sqrt{S}\gg p_T$, such final states are produced by the small-$x$ partons, and the resummation of $\log(1/x)$-enhanced
  terms into the unintegrated PDF (unPDF) can be implemented~\cite{CCFM}. Clearly, the regions of applicability of the TMD and $k_T$-factorization
  are overlapping, and they should match when $x \to 1$.

  Turning back to the photon pair production, we can conclude, that neither TMD nor $k_T$-factorization covers all available range
   of experimental data. Most of the cross section comes from the region where the diphoton has small $p_T$ and photons fly nearly back-to-back
    in the transverse plane, so additional QCD radiation is kinematically constrained to be soft and collinear, and the approach of SCET
     factorization will be preferable. On the contrary, at high $p_T$ and small $\Delta\phi$, the $k_T$-factorization will do a good job,
     as we will show below.

  The previous attempts to study the prompt diphoton production in $k_T$-factorization~\cite{Saleev_gg, A_Lipatov_gg} had their own problems.
   In the Ref.~\cite{Saleev_gg}, only LO $2\to 2$ subprocesses where taken into account. While, the PRA was used to obtain the gauge invariant
    expression for the $Q\bar{Q}\to \gamma\gamma$ matrix element with off-shell initial state Reggeized quarks ($Q$), the matrix
    element for the $RR\to \gamma\gamma$  with off-shell Reggeized gluons
     in the initial state  was taken the same as in the CPM. In fact, this contribution was overestimated in Ref.~\cite{Saleev_gg} due to the erroneous overall factor 4 in the partonic cross section of the subprocess $gg\to \gamma\gamma$ presented in the Ref.~\cite{BergerBraatenField}, which have lead to the accidental agreement with the early Tevatron data~\cite{CDF_old}. This factor was carefully checked against the results presented in the literature~\cite{BoxCPM,Bern_gg_gaga}, as well as by our independent calculations of the exact $RR\to\gamma\gamma$ amplitude, described in the Sec.~\ref{sec:box} of the present paper.

  In the Ref.~\cite{A_Lipatov_gg}, the attempt to take into account the NLO $2\to 3$ subprocesses was made, but manifestly non
  gauge invariant matrix elements where used both for $2\to 2$ and $2\to 3$ subprocesses. Also, the unavoidable double counting
  of additional real radiations between NLO $q^\star g^\star\to q\gamma \gamma$ subprocess and the unPDF was not subtracted,
  which have lead to the questionable conclusion, that no resummation of the effects of soft radiation is needed in the small-$p_T$ region to
  describe the data.

  In view of above mentioned shortcomings of the previous calculations, the present study has two main goals. The first one is
  to calculate the real part of NLO corrections to the process under consideration in the PRA, and develop the technique of subtraction
  of double counting between real NLO corrections and unPDF in PRA. The second goal is to calculate the matrix element of the quark-box
  subprocess $RR\to \gamma\gamma$ in PRA, taking into account the exact dependence on the transverse momenta of initial-state Reggeized gluons.

  The present paper has the following structure, in Sec.~\ref{sec:basicPRA} the relevant basics of the PRA formalism are outlined and
  the amplitude for the LO subprocess $Q\bar{Q}\to \gamma\gamma$ is derived. In the Sec.~\ref{sec:realNLO} the NLO $2\to 3$ amplitudes
  are derived and the procedure for the subtraction of double counting between NLO real corrections and unPDF is explained.
  In the Sec.~\ref{sec:box} the computation of the amplitude for the quark-box subprocess $RR\to \gamma\gamma$ is reviewed,
  and in the Sec.~\ref{sec:num} we compare our numerical results with the most recent CDF~\cite{CDF_data_2012} and ATLAS~\cite{ATLAS_data_2013} data. Our conclusions are
  summarized in the Sec.~\ref{sec:conclusions}.

\section{Basic formalism and LO contribution}
\label{sec:basicPRA}

  As collinear factorization is based on the property of factorization of collinear singularities in QCD~\cite{CollFact},
  the $k_T$-factorization is based on the Balitsky-Fadin-Kuraev-Lipatov (BFKL)~\cite{BFKL} (see~\cite{QMRKrev, IFL_book} for the review)
  factorization of QCD amplitudes in the Multi-Regge Kinematics (MRK), i. e. in the limit of the high scattering energy and fixed
  momentum transfers. For example, the amplitude for the subprocess $q(q_1)+q(q_2)\to q(q_3)+ g(q_4)+ q(q_5)$ in the limit when
  \begin{equation*}
  s_{34} \gg -t_{13},\ s_{45} \gg -t_{25},
  \end{equation*}
  where $s_{ij}=(q_i+q_j)^2$, $t_{ij}=(q_i-q_j)^2$, has the form of the amplitude with exchange of the effective Reggeized particle in the $t$-channel:
  \begin{equation}
  {\cal A}^{c,\mu}=2s \left(\bar{u}(q_3)\gamma^{r_1}u(q_1)\right) \cdot\frac{1}{t_{13}}\left(
\frac{s_{34}}{s_0}\right)^{\omega(t_{13})} \cdot \Gamma_{r_1
r_2}^{c, \mu} (q_{t1},q_{t2}) \cdot \frac{1}{t_{25}}\left(\frac{s_{45}}{s_0}\right)^{\omega(t_{25})}\cdot \left(\bar{u}(q_5)\gamma^{r_{2}} u(q_2)\right),\label{ampl_MRK}
  \end{equation}
  where $q_{t1}=q_1-q_3$, $q_{t2}=q_5-q_2$, $c,\ r_1,\ r_2$ are the color indices, $\gamma^r$ is the effective $qqR$ vertex,
  $\Gamma^{c,\mu}_{r_1,r_2}(q_{t1},q_{t2})$ is the central gluon production vertex $RRg$, $\omega(t)$ is the gluon Regge trajectory.
  The Slavnov-Taylor identity $ (q_{t1}-q_{t2})_\mu\Gamma_{r_1 r_2}^{c,\mu}(q_{t1},q_{t2})=0$ holds for the effective production vertex,
  which ensures the gauge invariance of the amplitude.

  The analogous form for the MRK asymptotics of the amplitude with quark exchange in the $t$-channel was shown to hold in the
  Leading Logarithmic Approximation (LLA) in~\cite{FadinSherman}. For the review of modern status of the quark Reggeization
  in QCD see Ref.~\cite{BogFad}.

  The Regge factor $s^{\omega(t)}$ resums the loop corrections enhanced by the $\log(s)$ to all orders in strong coupling constant $\alpha_s$,
  and the dependence on the arbitrary scale $s_0$ should be canceled by the analogous dependence of the effective vertices,
  taken in all orders of perturbation theory. The Reggeized gluon in the $t$-channel is a scalar particle in the adjoint
  representation of the $SU(N_c)$. In the MRK limit, when all three particles in the final state are highly separated in rapidity,
  the light-cone momentum components carried by the Reggeons in $t$-channels obey the hierarchy $q_t^\mp\ll |{\bf q}_{t\perp}|\sim q_t^\pm$,
  so in the strict MRK limit, the ``small'' light-cone component is usually neglected.

\begin{turnpage}
\begin{figure}[H]
\includegraphics[scale=1]{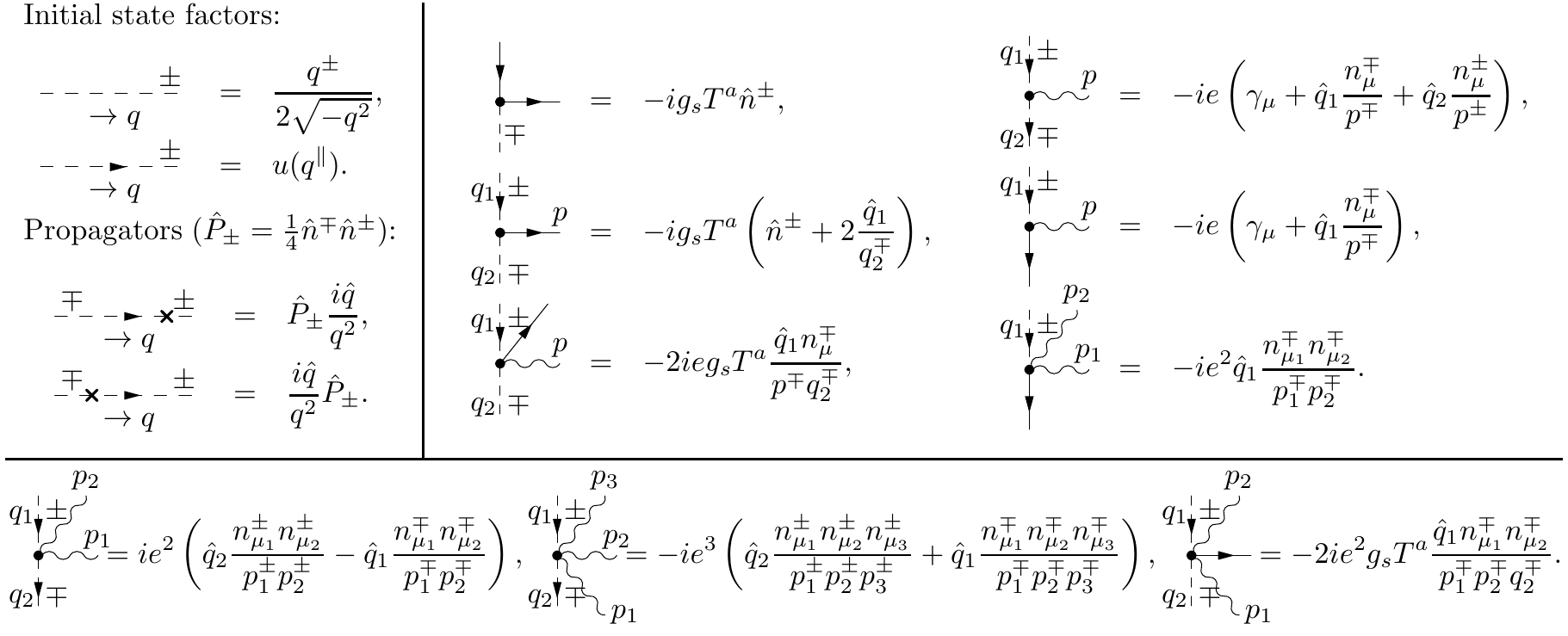}
\caption{\label{fig:FRs} The Feynman Rules of the effective theory~\cite{LipatovEFT, LipVyaz},
relevant for the present study. The propagators, factors, corresponding to the Reggeized quarks
(dashed lines with arrows) and gluons (dashed lines) in the initial state of the hard subprocess
and necessary interaction vertices are presented.  All momenta for the vertices are incoming. }
\end{figure}
\end{turnpage}

  To go beyond the LLA in $\log(s)$, one needs to consider the processes with a few clusters of particles
  in the final state, which are highly separated in rapidity, but keeping the exact kinematics within clusters. This is so called
  Quasi-Multi-Regge Kinematics(QMRK), and to obtain the amplitudes in this limit, the gauge invariant effective action for high energy processes
  in QCD was introduced in the Ref.~\cite{LipatovEFT}. Apart from the usual quark and gluon fields of QCD, which are supposed to live within
  a fixed rapidity interval, the fields of Reggeized gluons~\cite{LipatovEFT} and Reggeized quarks~\cite{LipVyaz} are introduced to communicate
  between the different rapidity intervals. To keep the $t$-channel factorized form of the amplitudes in the QMRK limit, the Reggeon fields
  have to be gauge invariant, which leads to the specific form of their nonlocal interaction with the usual QCD fields, containing the Wilson lines.
  Gauge invariance of Reggeon fields also ensures the gauge invariance of the effective emission vertices, which describe the production of particles
  within the given interval of rapidity, as it was the case in~(\ref{ampl_MRK}). The Feynman Rules (FRs) of the effective theory are collected
  in the refs.~\cite{FeynRules, LipVyaz}, but for the reader's convenience, we also list the FRs relevant for the purposes
  of the present study in the Fig.~\ref{fig:FRs}. To compute the hard-scattering matrix elements in PRA one have to combine the
  FRs of the Fig.~\ref{fig:FRs} with the usual FRs of QCD and QED, and use the factors for the Reggeons in the
  initial-state of the hard subprocess, also defined in the Fig.~\ref{fig:FRs} to be compatible with the normalization of the unPDF described below.

  Recently, the new scheme to obtain gauge-invariant matrix elements for $k_T$ factorization, by exploiting the spinor-helicity representation and recursion relations for the tree-level amplitudes was introduced~\cite{AVH:2013, AVH:2014}. This technique is equivalent to the PRA for the tree-level amplitudes without internal Reggeon propagators, however, the construction of the subtraction terms in the Sec.~\ref{sec:realNLO} requires the usage of the FRs of the refs.~\cite{FeynRules, LipVyaz}.

  So far as, the form of the central production vertex, propagators of Reggeized gluons and Regge trajectories, depends only on the
  quantum numbers of the Reggeon in the $t$-channel, and do not depend on what particles are in the initial state, the cross section
  of the production of particles to the central rapidity region in the inelastic $pp$ collisions can be written in the form~\cite{kTf}:
  \begin{equation}
  d\sigma=\sum\limits_{i,j}\int \frac{dx_1}{x_1}\int\frac{d^2 {\bf q}_{T1}}{\pi} \Phi_i(x_1,t_1,\mu_F^2) \int \frac{dx_2}{x_2}\int\frac{d^2 {\bf q}_{T2}}{\pi} \Phi_j(x_2,t_2,\mu_F^2) d\hat{\sigma}_{ij}(q_1, q_2),\label{eqn:kTfact}
  \end{equation}
  where the sums are taken over the parton species, $q_{1,2}=x_{1,2}P_{1,2}+q_{T1,2}$ are the momenta of the partons,
  $q_{1,2}^2=-{\bf q}_{T1,2}^2=-t_{1,2}$, $P_{1,2}$  are the four-momenta of the protons, $2P_1P_2=S$, $d\hat{\sigma}$ is
  the partonic cross section with Reggeized partons in the initial state. In what follows we will often use the Sudakov decomposition
  of the momenta:
\begin{equation*}
  k=\frac{1}{2}\left(n^+ k^-+ n^- k^+\right)+ k_T,
\end{equation*}
  where $k^\pm=k^0\pm k^3$, $P_1=\sqrt{S} n^-/2$, $P_2=\sqrt{S}n^+/2$, $(n^+)^2=(n^-)^2=0$, $n^+n^-=2$.

   The unPDF $\Phi_i(x, {\bf q}_T^2,\mu_F^2)$ is unintegrated over the transverse momentum ${\bf q}_T^2$, but still integrated over
   the "small" light-cone component of momentum, so this light-cone component is neglected in the hard scattering part,
   which is hence {\it formally} in the QMRK with the ISR and therefore it is gauge invariant. The exact kinematics will be restored
   by the higher-order QMRK corrections. The factorization scale $\mu_F^2$ is introduced to keep track of the position of the
   hard process on the axis of rapidity.

  The unPDF is normalized on the CPM number density PDF via:
  \begin{equation*}
  \int\limits^{\mu_F^2} dt \Phi_i(x,t,\mu_F^2) = x f_i(x,\mu_F^2).
  \end{equation*}

  In the case of an inelastic scattering of objects with intrinsic hard scale, such as photons with high center-of-mass energy and virtuality,
  the evolution of the unPDFs is governed by the large $\log(1/x)$ and they satisfy the BFKL evolution equation~\cite{BFKL}.
In proton-proton collisions, the initial state does not provide us
with intrinsic hard scale, therefore, the $k_T$-ordered
Dokshitzer-Gribov-Lipatov-Altarelli-Parisi (DGLAP)~\cite{DGLAP}
evolution at small $k_T$ should be merged with rapidity-ordered BFKL
evolution at high-$k_T$ final steps of the ISR cascade.

 The last problem is highly nontrivial and equivalent to the complete resummation of the $\log(k_T)$-enchanced terms in the BFKL kernel.
 A few phenomenological schemes to compute unPDFs of a proton where proposed, such as the Ciafaloni-Catani-Fiorani-Marchesini (CCFM)
 approach~\cite{CCFM}, the Bl\"umlein approach~\cite{Blumlein} and the Kimber-Martin-Ryskin approach~\cite{KMR}. In the LO calculations in PRA,
 the definition of the hard-scattering coefficient $d\hat{\sigma}$ is independent on the approximations made in the unPDF, so any unPDF
 can be used, and spread between them gives the theoretical uncertainty. In fact, the recent studies~\cite{OpenCharm, Bmesons} show,
 that in the realistic kinematical conditions, the LO calculations with KMR and recent version of the CCFM unPDFs~\cite{TMDLIB} give
 very close results. At the NLO, one should develop the proper matching scheme between the unPDF and corrections included into
 the hard-scattering kernel, which introduces a difference in treatment of different unPDFs.

  In the present paper we will work with the version of the KMR formula for the unPDFs, described in the Ref.~\cite{KMR_NLO}.
  The KMR prescription introduces the simplest possible scenario, where the $k_T$-ordered DGLAP chain of the emissions is followed
  by exactly one emission, ordered in rapidity with the particles produced in the hard subprocess. Due to the strong $k_T$-ordering
  of the DGLAP evolution, the transverse momentum of the parton in the initial-state of the hard subprocess is approximated to come
  completely from the last step of the evolution. With this approximations, one can obtain the unPDF from the conventional
  collinear PDF as follows:

  \begin{equation}
  \Phi_i(x,q_T^2,\mu^2)=\frac{1}{q_T^2}\int\limits_x^1 dz T_i(q^2,\mu^2)  \frac{\alpha_s(q^2)}{2\pi} \sum\limits_{j} P_{ij}(z) f_j\left(\frac{x}{z},q^2 \right)\theta\left(\Delta_{ij}(q_T^2,\mu^2)-z\right), \label{eqn:KMR}
  \end{equation}
  where $P_{ij}(z)$- DGLAP splitting function, $q^2=q_T^2/(1-z)$ -- virtuality of the parton in the $t$-channel, the Sudakov formfactor $T_i$ is defined as:

  \begin{eqnarray}
  T_i(q^2,\mu^2)=\exp\left\lbrace -\int\limits_{q^2}^{\mu^2} \frac{dk^2}{k^2} \frac{\alpha_s(k^2)}{2\pi} \sum\limits_{i,j}\int\limits_0^{1} d\xi \xi P_{ij}(\xi) \theta\left(\Delta_{ij}(k^2(1-\xi),\mu^2)-\xi\right) \right\rbrace,\label{eqn:Sud}
  \end{eqnarray}
  and the ordering in rapidity between the last parton emission and the particles produced in the hard subprocess is implemented via the following infrared cutoff~\footnote{The singularity of the $P_{gg}(z)$ splitting function at $z\to 0$ is also regularized by the cutoff $\theta(z-1+\Delta_{gg})$ in (\ref{eqn:KMR}) and (\ref{eqn:Sud}), which is not shown there for brevity. See the Ref.~\cite{KMR_NLO} for the details.}:
  \begin{equation*}
  \Delta_{ij}(q_T^2,\mu^2)=\frac{\mu}{\mu+q_T}\delta_{ij}+(1-\delta_{ij}).
  \end{equation*}

  In the present study, we use the version of the KMR formula~(\ref{eqn:KMR}) with LO DGLAP splitting functions, but NLO PDFs as a collinear input,
   because, as it was shown in the Ref.~\cite{KMR_NLO} the usage of the NLO PDFs and the exact scale $q^2$ are the most numerically important
   effects distinguishing the LO KMR distribution of the Ref.~\cite{KMR} and the NLO prescription of the Ref.~\cite{KMR_NLO}. Also,
   as it will be shown in the Sec.~\ref{sec:realNLO}, the usage of the LO DGLAP splitting functions is compatible with the PRA, while at the NLO, the splitting functions should be recalculated using the effective theory
   of the refs.~\cite{LipatovEFT, LipVyaz}.

  The effects of the Sudakov resummation are known to be dominant in the doubly-asymptotic region, $q_T\ll \mu$ and $z\to\Delta$, which
  is most important in $pp$ collisions, therefore we use the Sudakov form-factor in~(\ref{eqn:Sud}). The opposite limit, when $z\to 0$
  and $q_T\ll \mu$, is captured by the Regge factor $s^{\omega(t)}$, but the proper procedure of matching of the double-logarithmic corrections,
  between Sudakov and Regge factors is also beyond the scope of the present study.

  Now we are at step to discuss the LO and NLO contributions to the prompt photon pair production in the PRA.
  There is only one LO ($O(\alpha^2\alpha_s^0)$) subprocess:
  \begin{eqnarray}
  Q(q_1)+\bar{Q}(q_2)\to \gamma(q_3)+\gamma(q_4)\label{proc:QQgaga}
  \end{eqnarray}

  \begin{figure}[H]
\includegraphics[scale=1]{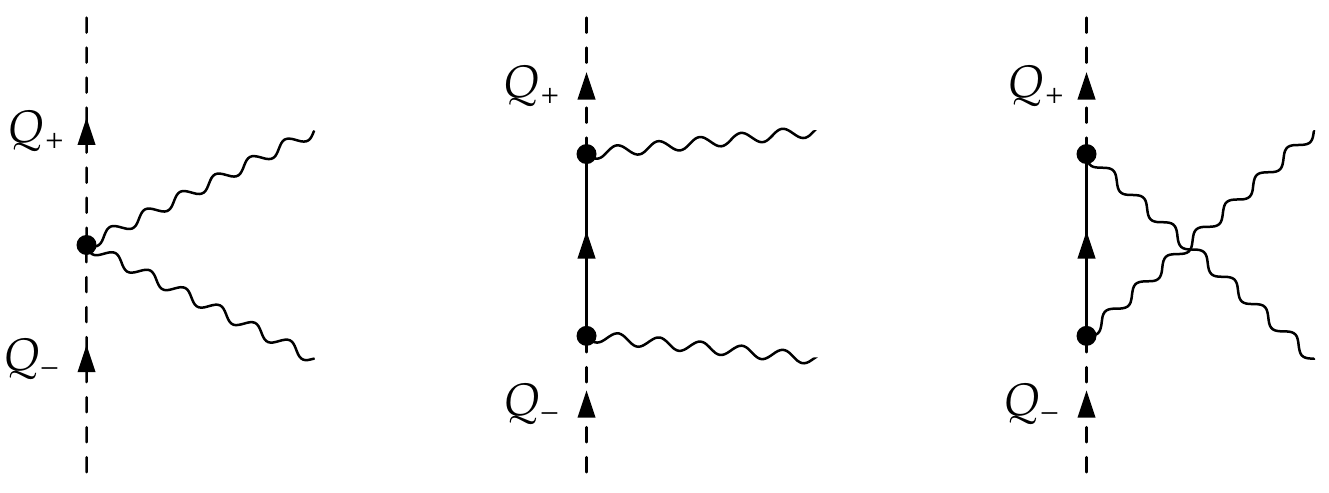}
\caption{\label{fig:QQgaga}  The set of Feynman diagrams for the LO subprocess~(\ref{proc:QQgaga}).}
\end{figure}

   The set of Feynman diagrams for this subprocess is presented on the Fig.~\ref{fig:QQgaga}. The amplitude of the process
    (\ref{proc:QQgaga}) obeys the Ward identity of Quantum Electrodynamics (QED)\emph{}, and the amplitude squared and averaged over the spin and color quantum numbers
    of the initial state, which was obtained in a first time in the Ref.~\cite{Saleev_gg}, has the form:
   \begin{eqnarray}
\overline{|{\cal A}(Q \bar Q\to \gamma \gamma)|^2}&=&
\frac{32}{3}\pi^2e_q^4\alpha^2\frac{x_1x_2}{a_3a_4b_3b_4S\hat t\hat
u}\Bigl(w_0+w_1S+w_2S^2+w_3S^3\Bigr),\label{amp:QQ}
   \end{eqnarray}
    where $a_3=q_3^+/\sqrt{S}$, $a_4=q_4^+/\sqrt{S}$, $b_3=q_3^-/\sqrt{S}$, $b_4=q_4^-/\sqrt{S}$, $\hat s=(q_1+q_2)^2$, $\hat t=(q_1-q_3)^2$, $\hat
u=(q_1-q_4)^2$, $x_1=a_3+a_4$, $x_2=b_3+b_4$, $\alpha=e^2/(4\pi)$, $e_q$ is the electric charge of the quark in the units of the electron charge
and the coefficients $w_i$ can be represented as follows:
\begin{eqnarray}
w_0=t_1t_2(t_1+t_2)-\hat t\hat u(\hat t+\hat u),\nonumber
\end{eqnarray}
\begin{eqnarray}
-w_1&=&t_1t_2(a_3-a_4)(b_3-b_4)+t_2x_1(b_4\hat t+b_3\hat u)+\nonumber\\
&+&t_1x_2(a_3\hat t+a_4\hat u)+\hat t\hat
u(a_3b_3+2a_4b_3+2a_3b_4+a_4b_4),\nonumber
\end{eqnarray}
\begin{eqnarray}
-w_2=b_3b_4x_1^2t_2+a_3a_4x_2^2t_1+a_3b_4\hat
t(x_2a_3+a_4b_4)+a_4b_3\hat u(a_3b_3+a_4x_2),\nonumber
\end{eqnarray}
\begin{eqnarray}
-w_3=a_3a_4b_3b_4\Bigl(a_3b_4\Bigl(\frac{\hat t}{\hat
u}\Bigr)+a_4b_3\Bigl(\frac{\hat u}{\hat t}\Bigr)\Bigr).\nonumber
\end{eqnarray}
Taking the collinear limit $t_{1,2}\to 0$ of~(\ref{amp:QQ}), one can
reproduce the standard CPM result for the amplitude
   $q\bar{q}\to \gamma\gamma$:

  \begin{equation*}
  \overline{|{\cal A}(q \bar q\to \gamma \gamma)|^2} = \frac{32}{3}\pi^2 e_q^4 \alpha^2 \left( \frac{\hat{t}}{\hat{u}} +
  \frac{\hat{u}}{\hat{t}} \right).
  \end{equation*}
  In the next section we will discuss the tree-level NLO corrections.

\section{Real NLO corrections}
\label{sec:realNLO}

  The tree-level NLO ($O(\alpha^2\alpha_s^1)$) subprocesses are:
  \begin{eqnarray}
  Q(q_1)+R(q_2)\to \gamma(q_3)+\gamma(q_4)+q(q_5), \label{proc:QRgagaq}\\
  Q(q_1)+\bar{Q}(q_2)\to \gamma(q_3)+\gamma(q_4)+g(q_5). \label{proc:QQgagag}
  \end{eqnarray}

\begin{figure}[H]
\includegraphics[scale=1]{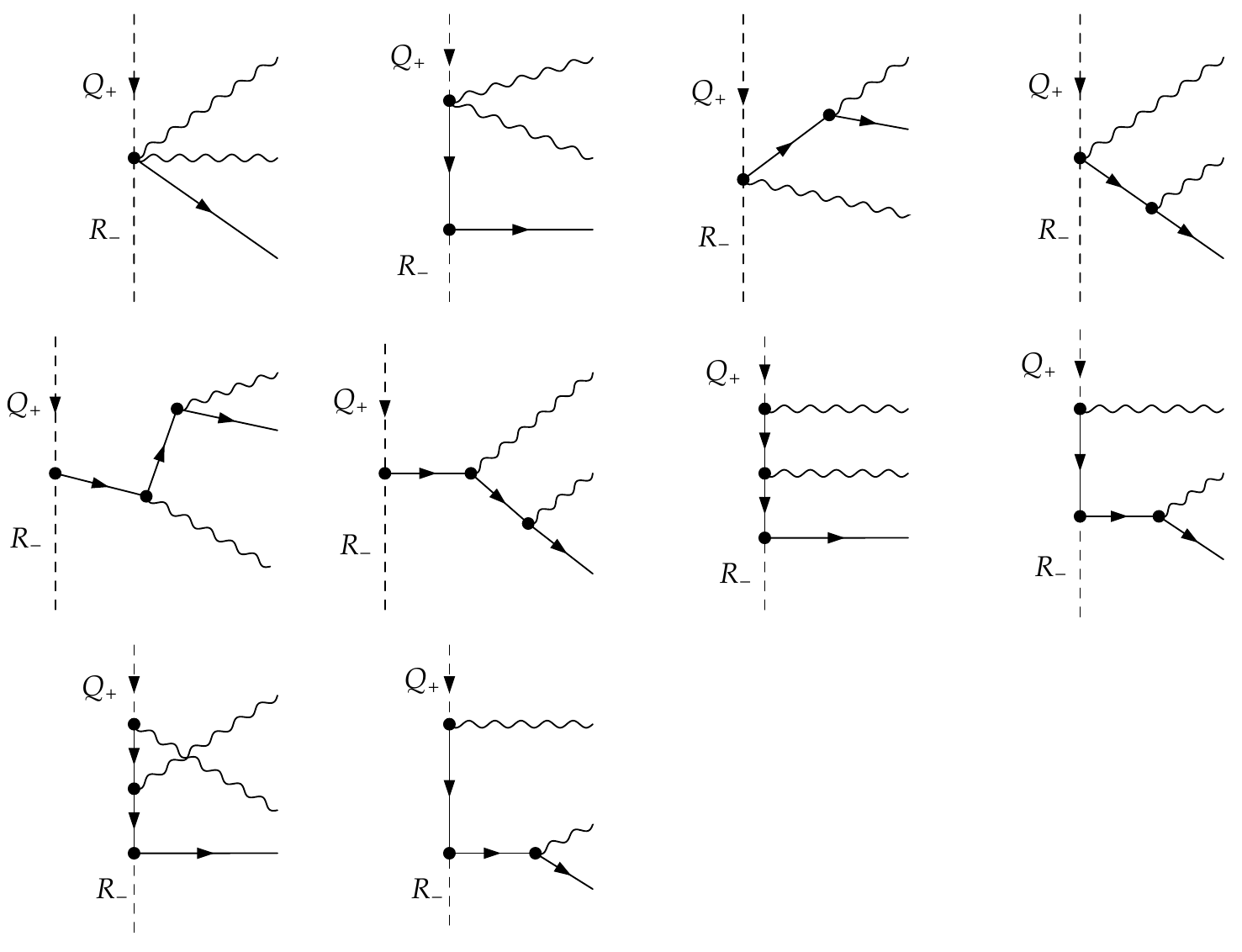}
\caption{\label{fig:QRqgaga} The set of the Feynman diagrams for the NLO subprocess~(\ref{proc:QRgagaq}).}
\end{figure}

\begin{figure}[H]
\includegraphics[scale=1]{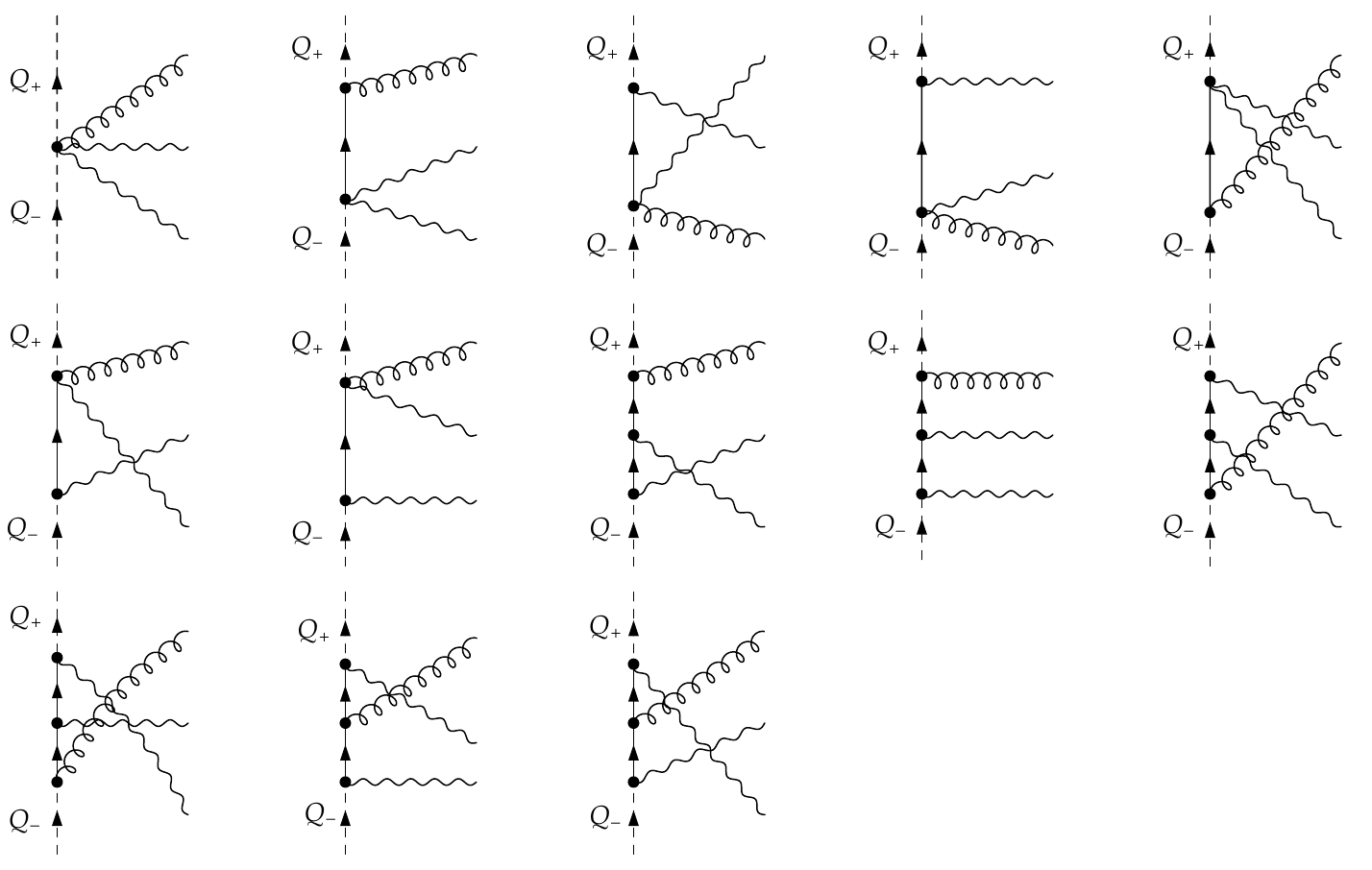}
\caption{\label{fig:QQgagag} The set of the Feynman diagrams for the NLO subprocess~(\ref{proc:QQgagag}). }
\end{figure}

  The sets of Feynman diagrams for them are presented in the Figs.~\ref{fig:QRqgaga} and~\ref{fig:QQgagag}.
  The FRs of the Fig.~\ref{fig:FRs} where implemented as the model file for the
  \texttt{FeynArts}~\cite{FeynArts},
  \texttt {Mathematica} based package, and the computation of the squared matrix elements was performed using
  \texttt{FeynArts},
  \texttt{FeynCalc}~\cite{FeynCalc}, and \texttt{FORM} programs.
   It was checked analytically, that the amplitudes for the NLO subprocesses (\ref{proc:QRgagaq}) and (\ref{proc:QQgagag}) obey the Ward (Slavnov-Taylor)
    identities with respect to all final state photons (gluons) independently on the transverse momentum of the initial state Reggeized partons. Unfortunately,
     the obtained expressions are too large and non-informative to present them here.

  The squared matrix element of the subprocess (\ref{proc:QRgagaq}) contains the collinear singularity, when the three-momentum
  of the quark becomes collinear to the three-momentum of one of the photons. This collinear singularity can be absorbed into
  the nonperturbative parton-to-photon fragmentation function, and then, the theoretical cross section is represented as the
  sum of {\it direct} contribution, where the collinear singularity is subtracted, according to e.~g. $\overline{MS}$ scheme,
  and {\it fragmentation} contribution, which is equal to the convolution of the cross section of the parton
  production in pQCD
 and the parton-to-photon fragmentation function.
   Experimental (hard-cone) isolation condition require the amount of hadronic energy within the photon isolation cone of
   the radius $R$ to be smaller than the fixed value $E_T^{(ISO)}\sim O(1)$ GeV:
  \begin{equation}
  E_T^{(had)}\left(r<R\right)<E_T^{(ISO)},\label{hcISO}
  \end{equation}
   where $r=\sqrt{\Delta\eta^2+\Delta\phi^2}$ is the distance in the pseudorapidity--azimuthal angle plane, $E_T^{(had)}(r<R)$
   is the amount of the hadronic transverse energy within the isolation cone around the photon. This isolation condition strongly
   suppresses the fragmentation component, but at high energies, fragmentation is still non-negligible, constituting up to
   the 20\% of the cross section~\cite{Diphox}.

  The proper treatment of the collinear singularity, considerably complicates the analytical computations both in the
  NNLO of CPM~\cite{2gNNLO, LHreport}, and in the NLO of PRA. Since in the PRA, the part of transverse momentum is provided
  to the hard subprocess by the phenomenological unPDFs. To avoid this difficulties, one can define the direct part of the
  cross section in the infrared-safe way, using the smooth-cone isolation condition~\cite{Frixione}:
  \begin{equation}
  r<R\Rightarrow E_T^{(had)}(r)<E_T^{(ISO)}\chi(r),\label{FrixISO}
  \end{equation}
  where $\chi(r)=\left(\frac{1-\cos(r)}{1-\cos(R)}\right)^n$, $n\ge 1/2$. The isolation condition (\ref{FrixISO}) is easy
  to implement into the process of Monte-Carlo integration, and it makes the cross section of the subprocess (\ref{proc:QRgagaq}) finite,
  because the collinear singularities associated with the initial-state are regularized by the unPDF. Applying the smooth-cone isolation
  to our calculation we are completely eliminating the need in the fragmentation component~\cite{Frixione, LHreport}, but of course this
  isolation do not match to the experimental one. However, as it was shown in the Ref.~\cite{LHreport}, the cross section obtained with
  the isolation condition (\ref{FrixISO}) is a lower estimate for the direct plus fragmentation cross section, obtained with the hard-cone
  isolation. Numerically, for $n=1$ this estimate is very good, since it reproduces the NLO results with standard isolation with the
accuracy of $O(1\%)$~\cite{LHreport}. Having in mind, that we are
going to discuss $O(50\%-100\%)$ NLO effects, we will apply the isolation
condition (\ref{FrixISO}) to our present calculations.

  The cross section of the subprocess (\ref{proc:QQgagag}) is also finite, because the Sudakov formfactor decreases
  in the region $q_T^2\ll \mu^2$ faster than any positive power of $q_T$ and therefore regularizes the collinear and
  soft singularities of the matrix element of the subprocess~(\ref{proc:QQgagag}) in the limit of $q_{T5}\to 0$.

  In the factorization formula (\ref{eqn:kTfact}), the part of the ISR, highly separated in rapidity from the particles produced
  in the hard subprocess is included into unPDFs, and the effects of the additional radiations close in rapidity to the hard subprocess,
  should be taken order by order in $\alpha_s$ in the hard-scattering coefficient. Therefore, the corresponding MRK asymptotics should
  be subtracted from the NLO QMRK contributions~(\ref{proc:QRgagaq}), and (\ref{proc:QQgagag}) to avoid the double counting, when the additional
  parton is highly separated in rapidity from the central region. The analogous procedure of the ``localization in rapidity'' of the QMRK contributions was proposed in the refs.~\cite{BartelsNLO, ASV_NLO}. To be compatible with our definition of the KMR unPDF~(\ref{eqn:KMR}),
  this subtraction term should interpolate smoothly between the strict MRK limit, when additional parton goes deeply forward or backward
   in rapidity with fixed transverse-momentum, and collinear factorization limit, when the initial-state partons are nearly on-shell and
   additional parton has a small transverse momentum but it's rapidity is arbitrary. Below, such subtraction term is constructed in close
   analogy with the High Energy Jets approach~\cite{HEJ}.

\begin{figure}[H]
\includegraphics[scale=1]{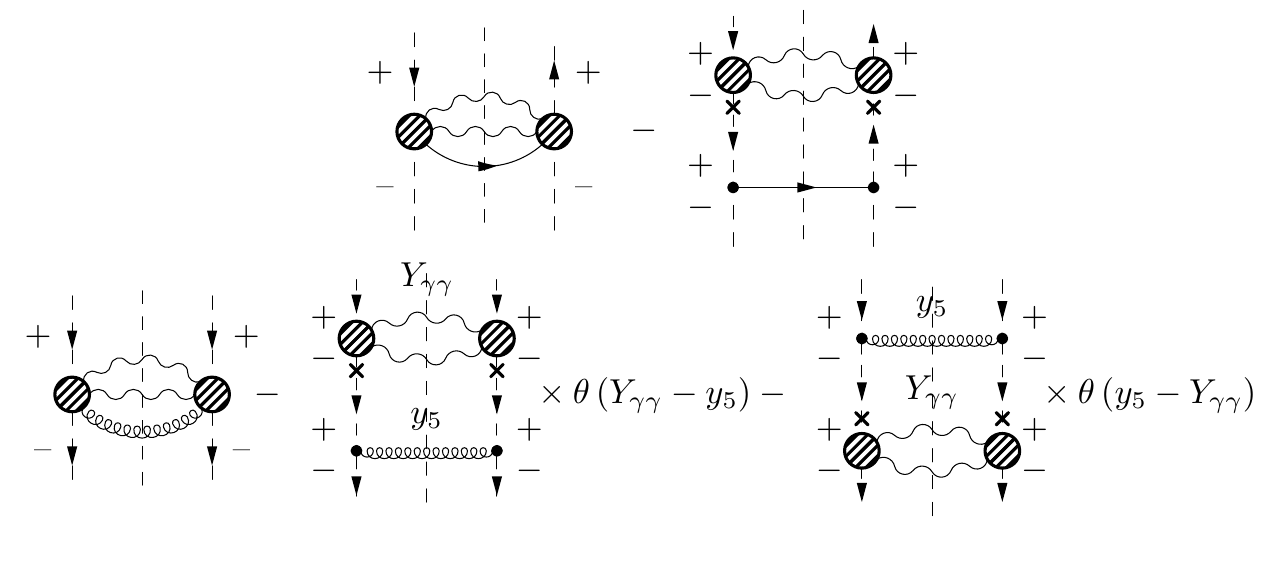}
\caption{\label{fig:subtr} Upper panel, the diagrammatic representation for the squared amplitude of the subprocess~(\ref{proc:QRgagaq}) and the corresponding mMRK subtraction term. Lower panel -- the same for the subprocess~(\ref{proc:QQgagag}). }
\end{figure}

  The Feynman diagrams for the subtraction terms, required for the squared amplitudes of the subprocesses~(\ref{proc:QRgagaq}),
  and (\ref{proc:QQgagag}) are shown in the Fig.~\ref{fig:subtr} and can be easily written according to the FRs of
  the Fig.~\ref{fig:FRs}. To extend the applicability of the subtraction terms outside of the strict MRK limit, one have to implement
  the exact $2\to 3$ kinematics for the subtraction terms, taking into account the exact $\hat{t}$-channel momentum in the propagator
  of the Reggeized quark. In what follows, we will refer to the amplitudes with the Reggeon propagators and vertices, but without
  kinematical approximations, characteristic for the MRK, as modified MRK (mMRK) amplitudes. As it was checked explicitly,
  the implementation of the exact kinematics do not destroy the gauge invariance of the subtraction terms with the Reggeized quarks
  in the $\hat{t}$-channels, presented in the Fig.~\ref{fig:subtr}, as it was the case for the mMRK amplitudes with the Reggeized gluons
  in the $\hat{t}$-channels in the Ref.~\cite{HEJ}.

\begin{figure}[H]
\includegraphics[scale=1]{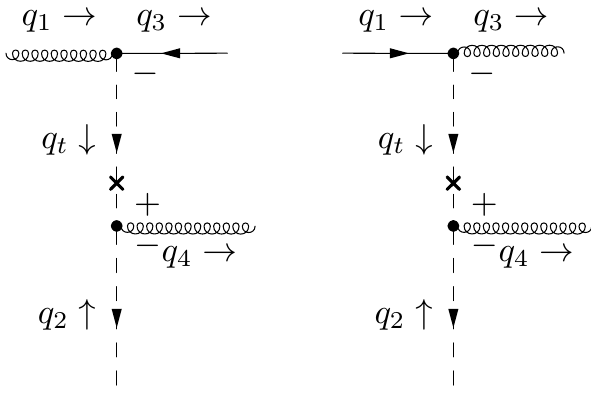}
\caption{\label{fig:Subp_ex} The Feynman diagrams for the mMRK subprodesses $g(q_1)+\bar{Q}(q_2)\to \bar{q}(q_3)+g(q_4)$ (left panel) and $q(q_1)+\bar{Q}(q_2)\to g(q_3)+g(q_4)$ (right panel).}
\end{figure}

  The last ambiguity, which we have to fix in the definition of our mMRK amplitudes is the position of the $\hat{P}_{\pm}$-projector in
  the numerator of the propagator of the Reggeized quark. In the MRK limit, the ``small'' light-cone component of the Reggeon momentum
  can be neglected and the projectors $\hat{P}_{\pm}$ commute with $\hat{q}_t$ under the sign of the trace, but outside of this limit,
  this is not true anymore. To fix this ambiguity, let's study the amplitudes of the mMRK subprocesses in the Fig.~\ref{fig:Subp_ex}.
  Explicitly, they have the forms:
  \begin{eqnarray}
  {\cal A}^{ab}_{\mu\nu}(g\bar{Q}\to \bar{q}g)&=& g_s^2\bar{v}(q_2^\parallel)\left(\gamma_\nu-\hat{q}_2\frac{n^+_\nu}{q_4^+}-\hat{q}_t\frac{n_\nu^-}{q_4^-}\right)T^b\hat{P}_+\frac{\hat{q}_t}{q_t^2}\left(\gamma_\mu-\hat{q}_t\frac{n_\mu^+}{q_1^+}\right)T^a v(q_3), \\
  {\cal A}^{ab}_{\mu\nu}(q\bar{Q}\to gg)&=&g_s^2\bar{v}(q_2^\parallel) \left(\gamma_\nu-\hat{q}_2\frac{n^+_\nu}{q_4^+}-\hat{q}_t\frac{n^-_\nu}{q_4^-}\right)T^b \hat{P}_+\frac{\hat{q}_t}{q_t^2}\left(\gamma_\mu+\hat{q}_t\frac{n^+_\mu}{q_3^+}\right)T^a u(q_1), \\
  {\cal A}^{a}_{\mu}(Q\bar{Q}\to g)&=& g_s \bar{v}(q_2^\parallel) \left(\gamma_\nu-\hat{q}_2\frac{n^+_\mu}{q_4^+}-\hat{q}_t\frac{n_\mu^-}{q_4^-}\right)T^a u(q_t^\parallel).
  \end{eqnarray}
  Taking the squared modulus of these amplitudes and averaging them over the initial-state spin and color quantum numbers, we get:
  \begin{eqnarray}
  \overline{|{\cal A}(g\bar{Q}\to \bar{q}g)|^2}&=&g_s^2\frac{\hat{s}+t_2}{\hat{s}+t_2+\hat{t}}\frac{2P_{qg}(z)}{z\hat{t}}\overline{|{\cal A}(Q\bar{Q}\to g)|^2},\label{eqn:A2gQ_qg}\\
  \overline{|{\cal A}(q\bar{Q}\to gg)|^2}&=&g_s^2\frac{\hat{s}+t_2}{\hat{s}+t_2+\hat{t}}\frac{2P_{qq}(z)}{z\hat{t}}\overline{|{\cal A}(Q\bar{Q}\to g)|^2},\label{eqn:A2qQ_gg} \\
  \overline{|{\cal A}(Q\bar{Q}\to g)|^2}&=&g_s^2\frac{C_AC_F}{N_c^2}({\bf q}_{t\perp}^2+t_2),\label{eqn:A2QQ_g}
  \end{eqnarray}
  where we have taken the limit $q_1^2=0$ to facilitate the study of the collinear singularity, $q_2^2=-t_2$,
  the invariants $\hat{s}$, $\hat{t}$, $\hat{u}$ are defined after the Eq.~(\ref{amp:QQ}), ${\bf q}_{t\perp}=({\bf q}_{T2}-{\bf q}_{T4})$,
  $z=1-{q_3^+}/{q_1^+}$ and $P_{qg}(z)=\frac{1}{2}(z^2+(1-z)^2)$, $P_{qq}(z)=C_F\frac{1+z^2}{1-z}$ are the LO DGLAP splitting functions.

  When $z\ll 1$ and $\hat{t}$-fixed, the partons 3 and 4 are in the MRK. In the opposite (collinear) limit $\hat{t}\to 0$, the squared amplitudes
  (\ref{eqn:A2gQ_qg}), and (\ref{eqn:A2qQ_gg}) factorize into the collinear singularity with the corresponding DGLAP splitting function
  and the squared amplitude (\ref{eqn:A2QQ_g}). From this example one can conclude, that the factor $\hat{q}_t$ should be taken together
  with the vertex of the MRK-emission to correctly reproduce the collinear singularity of the amplitude. This prescription is denoted by
  the crosses on the quark propagators in the Figs.~\ref{fig:FRs}, \ref{fig:subtr} and \ref{fig:Subp_ex}.

  The squared amplitudes (\ref{eqn:A2gQ_qg}, \ref{eqn:A2qQ_gg}) can also be used to explain the structure of the factorization
  formula~(\ref{eqn:kTfact}) and the unPDF~(\ref{eqn:KMR}). The presence of the exact DGLAP splitting functions in~(\ref{eqn:A2gQ_qg}), and
   (\ref{eqn:A2qQ_gg}) corresponds to the usage of the exact splitting functions in the unPDF~(\ref{eqn:KMR}).
   Factor $z$ in the denominators of~(\ref{eqn:A2gQ_qg}, \ref{eqn:A2qQ_gg}) is nothing but a flux factor of the $\hat{t}$-channel partons,
   which tells us, that for the Reggeized partons one should use the same flux factor $I=2Sx_1x_2$ as for the CPM partons.
   After the integration over the ``small'' light-cone component in the definition of the unPDF, the additional factor $1/(1-z)$ appears,
   which converts $\hat{t}$ into $\hat{t}(1-z)={\bf q}_{t\perp}^2$, that's why the $q_T^2$ and not $q^2$ stands in the
   denominator of (\ref{eqn:KMR}).

  The rapidity ordering conditions are imposed in the subtraction terms for the subprocess (\ref{proc:QQgagag})
  (see Fig.~\ref{fig:subtr}, lower panel), while for the case of the subprocess (\ref{proc:QRgagaq}) the rapidity of the quark in
  the final state is unconstrained (Fig.~\ref{fig:subtr}, upper panel). This corresponds to the fact, that in the KMR unPDF,
  the radiation of the gluon is ordered in rapidity with the particles, produced in the hard subprocess, while for the quark
  it is not the case. So, the mMRK terms constructed according to the Feynman diagrams in the Fig.~\ref{fig:subtr}
  are completely well-defined and correspond to the definition of the unPDF~(\ref{eqn:KMR}).

\section{The quark-box contribution}
\label{sec:box}
  The subprocess:
  \begin{equation}
  R(q_1)+R(q_2)\to \gamma(q_3)+\gamma(q_4),\label{proc:RRgaga}
  \end{equation}
   is described by the quark-box amplitude, and is formally NNLO ($O(\alpha^2\alpha_s^2)$), but it it's contribution to the total cross section
  is expected to be comparable with NLO contributions, due to the enhancement by two gluon unPDFs. The helicity amplitudes for
  the subprocess (\ref{proc:RRgaga}) could be written as:
  \begin{eqnarray}
  {\cal A}(RR,\lambda_3\lambda_4)=\frac{q_1^+q_2^-}{4\sqrt{t_1t_2}}n^-_{\mu_1} n^+_{\mu_2} \varepsilon^\star_{\mu_3}(\lambda_3) \varepsilon^\star_{\mu_4}(-\lambda_4){\cal M}^{\mu_1\mu_2\mu_3\mu_4}, \label{eqn:RRgaga}
  \end{eqnarray}
  where $\lambda_3$, $\lambda_4$ are the helicities of the photons in the final state and the fourth-rank vacuum polarization tensor has the form:
  \begin{eqnarray}
  {\cal M}^{\mu_1\mu_2\mu_3\mu_4}&=&2\int d^4q\left\lbrace\frac{\tr\left[(\hat{q}-\hat{q}_1)\gamma^{\mu_3}(\hat{q}+\hat{q}_2-\hat{q_4})\gamma^{\mu_4}(\hat{q}+\hat{q}_2)\gamma^{\mu_2}\hat{q}\gamma^{\mu_1}\right]}{(q-q_1)^2(q+q_2-q_4)^2(q+q_2)^2 q^2}\right.+\nonumber \\
  &+& \left.(q_3\leftrightarrow q_4,\mu_3\leftrightarrow \mu_4)+(q_2\leftrightarrow -q_4,\mu_2\leftrightarrow\mu_4)\right\rbrace , \label{eqn:4Mtensor}
  \end{eqnarray}
  where the factor $2$ takes into account the diagrams with the opposite direction of the fermion number flow.
  The following overall factor is taken out from the amplitude (\ref{eqn:RRgaga}):
  \begin{equation*}
  \frac{e^2g_s^2}{(2\pi)^4}\frac{\delta_{ab}}{2}\left(\sum\limits_q e_q^2\right).
  \end{equation*}
  We take the polarization vectors for the final-state photons in the form:
  \begin{equation*}
  \varepsilon^\mu(\lambda)=\frac{1}{\sqrt{2}}\left(n_x^\mu+i\lambda n_y^\mu\right),
  \end{equation*}
  where
  \begin{eqnarray*}
  n_x^\mu &=& \frac{1}{\Delta}\left( (q_3 q_4)q_2^\mu - (q_2q_4)q_3^\mu -(q_2q_3)q_4^\mu \right), \\
  n_y^\mu &=& -\frac{1}{\Delta}\epsilon^{\mu q_2 q_3 q_4},
  \end{eqnarray*}
  and  $\Delta=\sqrt{\hat{s}\hat{t}\hat{u}-\hat{s}t_1t_2}/2$.

  In the Ref.~\cite{KNS_photon_jet} we managed to obtain the compact result for the helicity amplitudes of the process
  $\gamma R\to \gamma g$, and explicitly demonstrate the cancellation of the spurious collinear singularity $1/(t_1 t_2)$ in the squared amplitude.
  For the process~(\ref{proc:RRgaga}) it turns out to be impossible to obtain the reasonably compact results, and the task is actually
  to obtain the representation for the helicity amplitudes which will be feasible for the numerical evaluation at all. To do this, we observe,
  that exploiting the Ward identity $q_{1,2}^{\mu_{1,2}} {\cal M}_{\mu_1\mu_2\mu_3\mu_4}=0$ for the tensor (\ref{eqn:4Mtensor}),
  one can make the following substitutions in (\ref{proc:RRgaga}):
  \begin{equation*}
  \frac{q_1^+q_2^-}{4\sqrt{t_1t_2}}(n^-)^{\mu_1}(n^+)^{\mu_2}\to n^{\mu_1}_{T1}n_{T2}^{\mu_2},
  \end{equation*}
  where $n_{T1,2}=q_{T1,2}/\sqrt{t_{1,2}}$.

  To get rid of the $\epsilon$-tensors and directly pass to the Passarino-Veltman reduction for the Feynman integrals with the scalar
  products in the numerator, we exploit the same trick as in Ref.~\cite{KNS_photon_jet}. We decompose the four-vectors $n_{T1,2}$ as follows:
  \begin{eqnarray}
  n_{T1}&=&\beta_0^{(1)} q_1 + \beta_3^{(1)} q_3 +\beta_4^{(1)} q_4 +\gamma_1 n_y,\\
  n_{T2}&=&\beta_0^{(2)} q_2 + \beta_3^{(2)} q_3 +\beta_4^{(2)} q_4 +\gamma_2 n_y,
  \end{eqnarray}
  and the vector $n_y$ is introduced via it's scalar products: $n_y^2=-1$, $n_yq_2=n_yq_3=n_yq_4=0$. The coefficients of this decomposition
  can be straightforwardly expressed through the Mandelstam invariants, transverse momenta of particles and azimuthal angles.
  After the Passarino-Veltman reduction, the helicity amplitudes where represented as a linear combinations of two, three and
  four-point scalar one loop integrals, and the cancellation of the Ultra-Violet (UV) and Infra-Red (IR) divergences was checked both analytically and numerically.
  The coefficients of this decomposition depends on 5 invariants $\hat{s},\ \hat{t},\ \hat{u},\ t_1,\ t_2$ and 8
  coefficients $\beta_i^{(j)}$, $\gamma_{1,2}$, i.e. 13 parameters in total. They can be represented as rational functions
  with tens of thousands  terms in the numerators. It turns out, that just to reliably check the cancellation of the UV and IR
  divergences numerically, one have to compute this coefficients with 30 digits of accuracy at least.

  Also, it was checked, both analytically and numerically, that the collinear limit for the squared helicity amplitudes (\ref{eqn:RRgaga}),
  defined as:
    \begin{equation*}
  \int\limits_0^{2\pi} \frac{d\phi_1 d\phi_2}{(2\pi)^2}\lim\limits_{t_{1,2}\to 0} |{\cal A}(RR,\lambda_3\lambda_4)|^2= \frac{1}{4} \sum\limits_{\lambda_{1,2}=\pm} |{\cal A}_{CPM}(\lambda_1\lambda_2,\lambda_3\lambda_4)|^2,
  \end{equation*}
  holds, where the $|{\cal A}_{CPM}(\lambda_1\lambda_2,\lambda_3\lambda_4)|^2$ is the squared helicity amplitude of the process
  $gg\to \gamma\gamma$ in the CPM. The numerical check of the collinear limit was performed by the technique described in the
  Ref.~\cite{KNS_photon_jet}. The numerical results for the subprocess~(\ref{proc:RRgaga}) will be presented in the next section,
  and the \texttt{FORTRAN} code for the calculation of the helicity amplitudes and differential cross sections of the
  process~(\ref{proc:RRgaga}), as well as for the $2\to 2$ subprocess (\ref{proc:QQgaga}) and $2\to 3$ subprocesses (\ref{proc:QRgagaq}), and
  (\ref{proc:QQgagag}) is available from authors on request.

\section{Numerical results}
\label{sec:num}

  The differential cross section of the $2\to 2$ subprocesses (\ref{proc:QQgaga}, \ref{proc:RRgaga}) can be represented as follows:
  \begin{equation}
  \frac{d\sigma}{dq_{T3}dq_{T4}d\Delta\phi dy_3 dy_4}=\frac{1}{2!}\int dt_1 \int\limits_0^{2\pi} d\phi_1 \sum\limits_{ij} \Phi_i(x_1,t_1,\mu_F^2) \Phi_j(x_2,t_2,\mu_F^2) \frac{q_{T3}q_{T4} \overline{|{\cal A}_{ij}|^2}}{2(2\pi)^3 (Sx_1x_2)^2}, \label{cs:2to2}
  \end{equation}
  where the factor $1/2!$ takes into account the identical nature of the photons, $q_{Ti}=|{\bf q}_{Ti}|$, $y_i$ are the rapidities of
  the final-state particles, $\Delta\phi$ is the azimuthal angle between transverse momenta of the photons, $\phi_1$ is the azimuthal angle
  between ${\bf q}_{T3}$ and ${\bf q}_{T1}$, $t_2=({\bf q}_{T3}+{\bf q}_{T4}-{\bf q}_{T1})^2$,
  $x_{1,2}=(q_{T3}e^{\pm y_3}+q_{T4}e^{\pm y_4})/\sqrt{S}$. The spectra differential in the diphoton invariant mass ($M$)
  and diphoton transverse momentum $p_T=\sqrt{q_{T3}^2+q_{T4}^2}$, could be obtained using the following substitutions:
  \begin{eqnarray}
  dM&=&\frac{q_{T3}C}{M} dq_{T4}, \label{eqn:subM}\\
  dp_T&=& \frac{D}{p_T} dq_{T4},\label{eqn:subpT}
  \end{eqnarray}
  where $C=\cosh(y_3-y_4)-\cos(\Delta\phi)$, $D=\sqrt{p_T^2-q_{T3}^2\sin^2(\Delta\phi)}$, $q_{T4}=M^2/(2q_{T3}C)$ for the case
  of~(\ref{eqn:subM}) and $q_{T4}=D-q_{T3}\cos(\Delta\phi)$ for the~(\ref{eqn:subpT}).

  For the $2\to 3$ subprocesses (\ref{proc:QRgagaq}, \ref{proc:QQgagag}), the formula for the differential cross section reads:
  \begin{eqnarray}
  \frac{d\sigma}{dq_{T3}dq_{T4}d\Delta\phi dy_3 dy_4 dy_5}=\frac{1}{2!}\int dt_1 \int\limits_0^{2\pi} d\phi_1 \int dt_2 \int\limits_0^{2\pi} d\phi_2 \times \nonumber\\
  \times \sum\limits_{ij} \Phi_i(x_1,t_1,\mu_F^2) \Phi_j(x_2,t_2,\mu_F^2) \frac{q_{T3}q_{T4} \overline{|{\cal A}_{ij}|^2}}{16(2\pi)^6 (Sx_1x_2)^2},\label{cs:2to3}
  \end{eqnarray}
  and the differential cross sections over the $p_T$ and $M$ could be obtained using the substitutions (\ref{eqn:subM}), and (\ref{eqn:subpT})
  as in the $2\to 2$ case.

\begin{figure}
  \includegraphics[width=\textwidth]{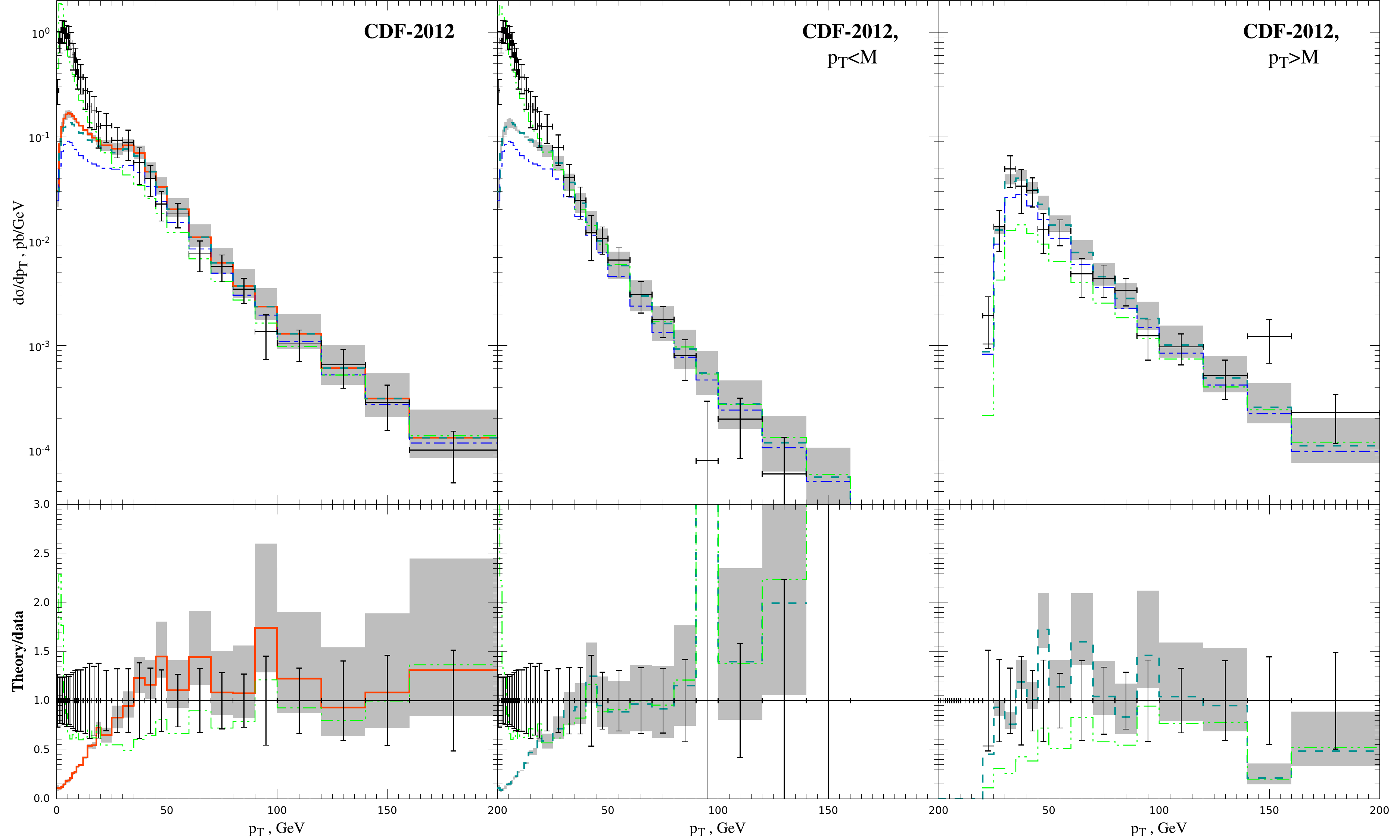}
  \caption{The $p_T$-spectra for the CDF-2012 dataset. The thick solid curve is the sum of the contributions (\ref{proc:QQgaga}),  (\ref{proc:QRgagaq})
  with mMRK subtraction, and (\ref{proc:RRgaga}). The thick dashed curve is the sum of the first two.
  The thin dash-dotted curve is the contribution of the subprocess (\ref{proc:QQgaga}) only. The thin dash-double-dotted curve is the corresponding \texttt{Diphox} (NLO CPM) prediction, taken from the Ref.~\cite{CDF_data_2012}, see the text for details.}\label{fig:CDF_pT}
  \end{figure}

\begin{figure}
  \includegraphics[width=0.45\textwidth]{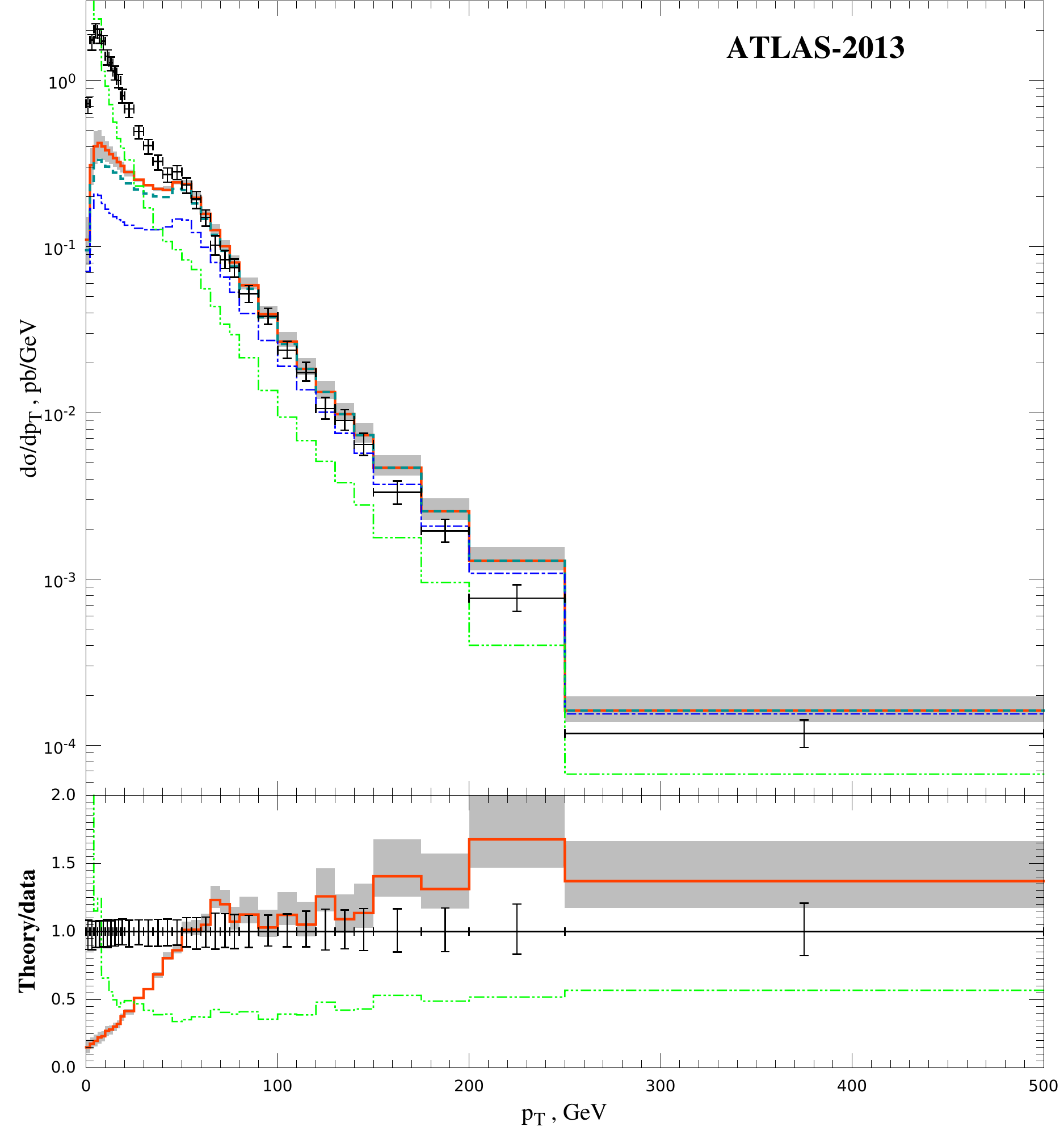}
   \includegraphics[width=0.45\textwidth]{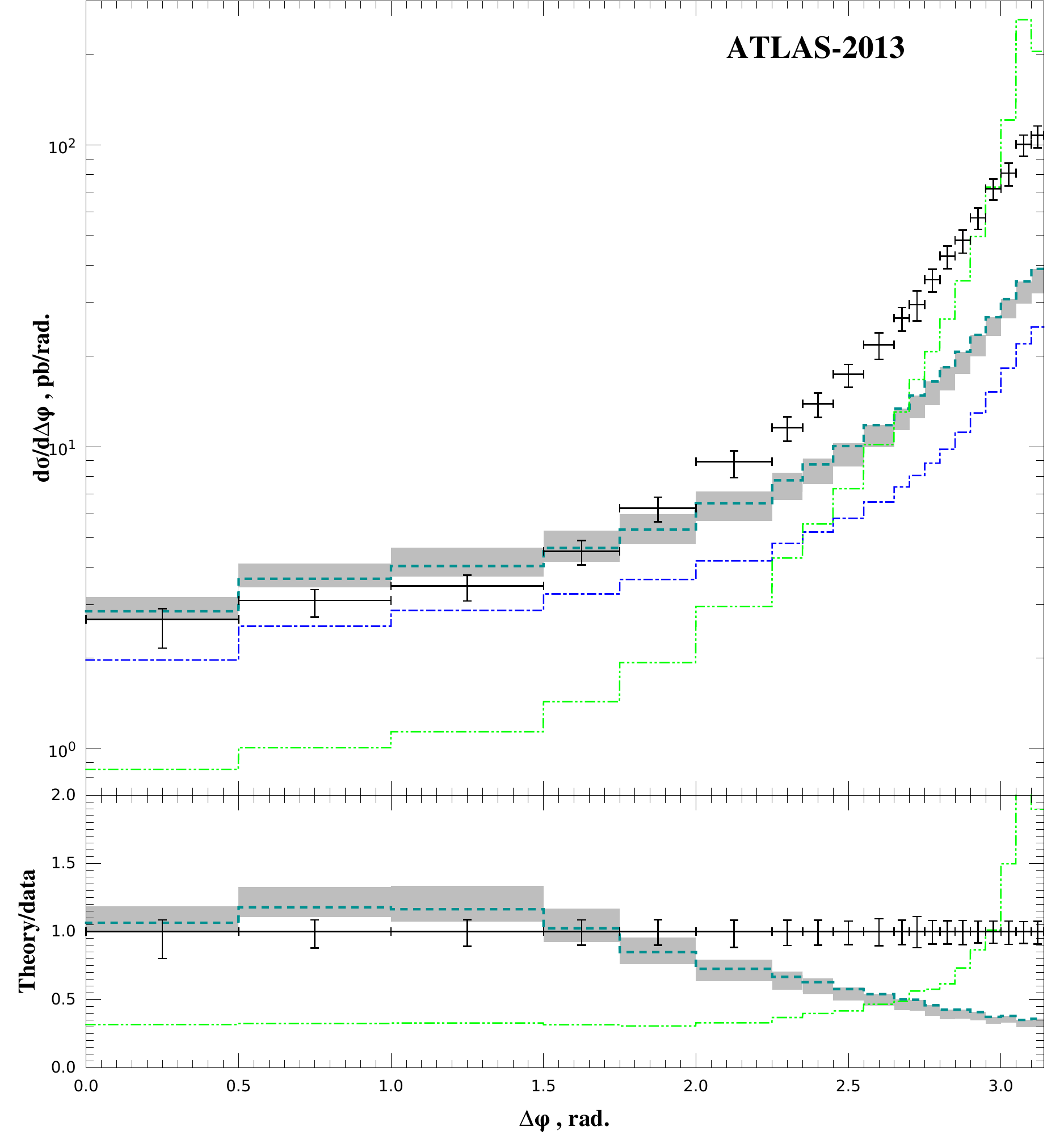}
  \caption{The $p_T$-spectra (left panel) and $\Delta\phi$-spectra (right panel) for the ATLAS-2013 dataset. The notation for the curves is the same as in the Fig.~\ref{fig:CDF_pT}, except the thin dash-double-dotted curve, which is the \texttt{Diphox} (NLO CPM) prediction, taken from the Ref.~\cite{ATLAS_data_2013}, see the text for details.}\label{fig:ATLAS_pT}
  \end{figure}

\begin{figure}
  \includegraphics[width=\textwidth]{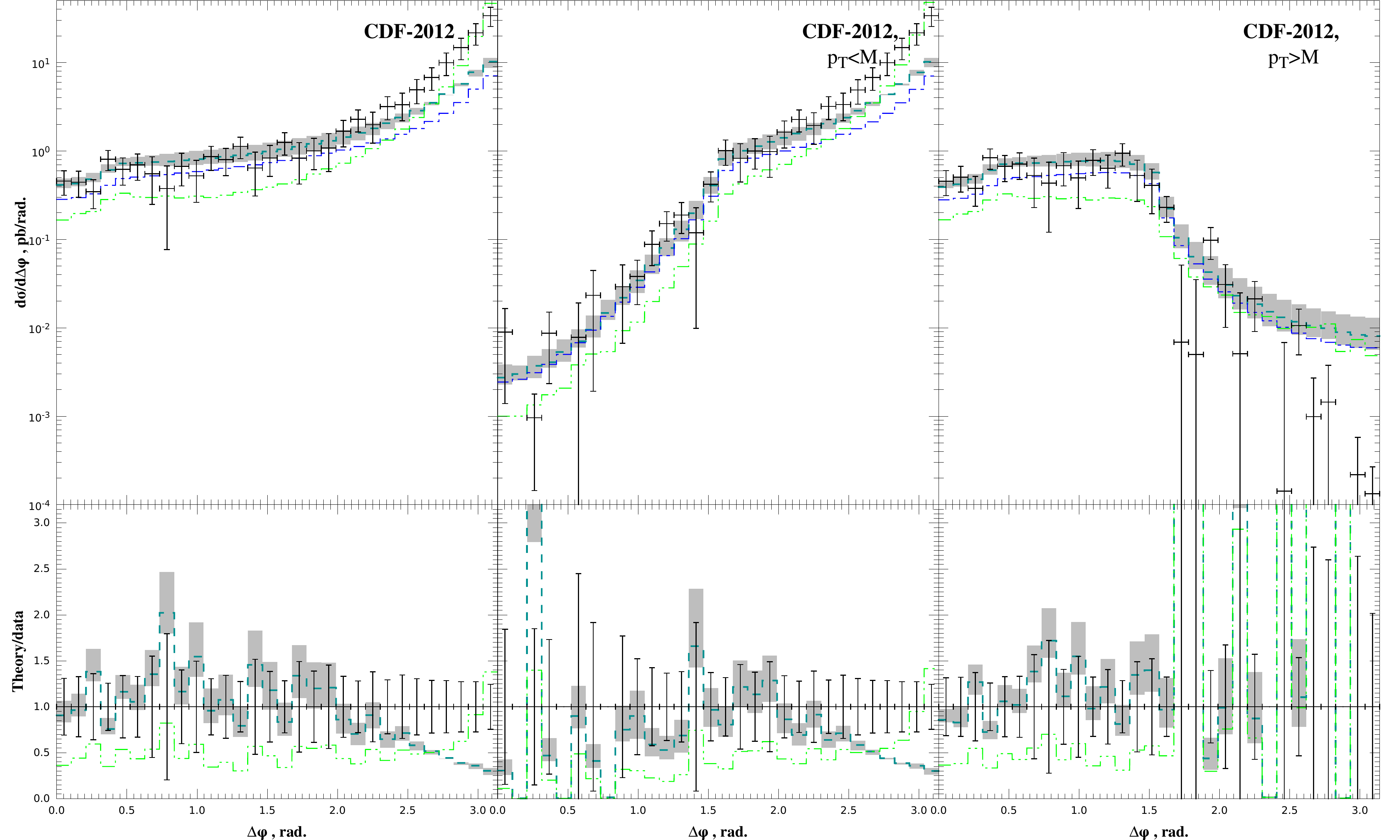}
  \caption{The $\Delta\phi$-spectra for the CDF-2012 dataset. The notation for the curves is the same as in the Fig.~\ref{fig:CDF_pT}. }\label{fig:CDF_PHI}
  \end{figure}

\begin{figure}
  \includegraphics[width=\textwidth]{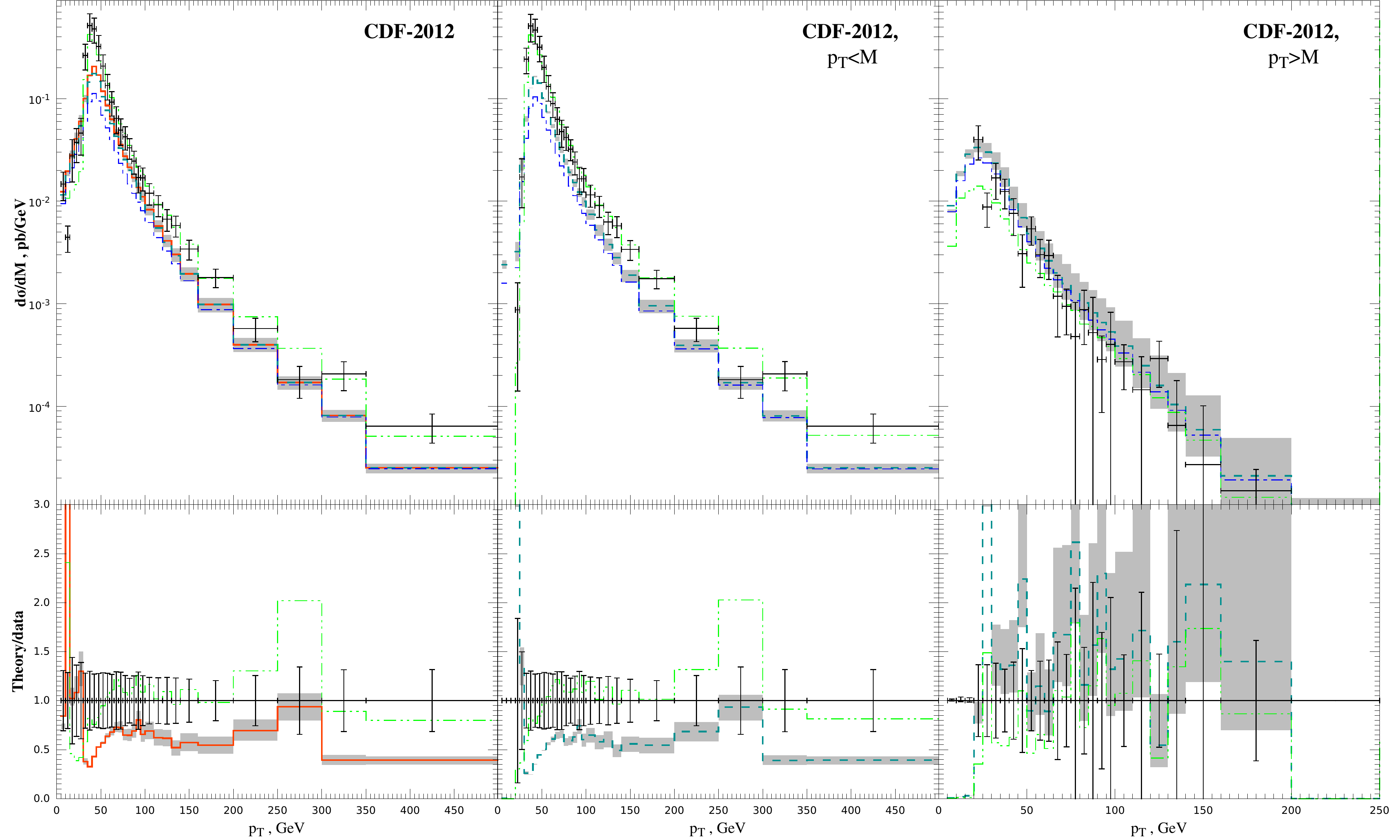}
  \caption{Diphoton invariant mass spectra for the CDF-2012 dataset. The notation for the curves is the same as in the Fig.~\ref{fig:CDF_pT}}\label{fig:CDF_M}
  \end{figure}

\begin{figure}
  \includegraphics[width=0.5\textwidth]{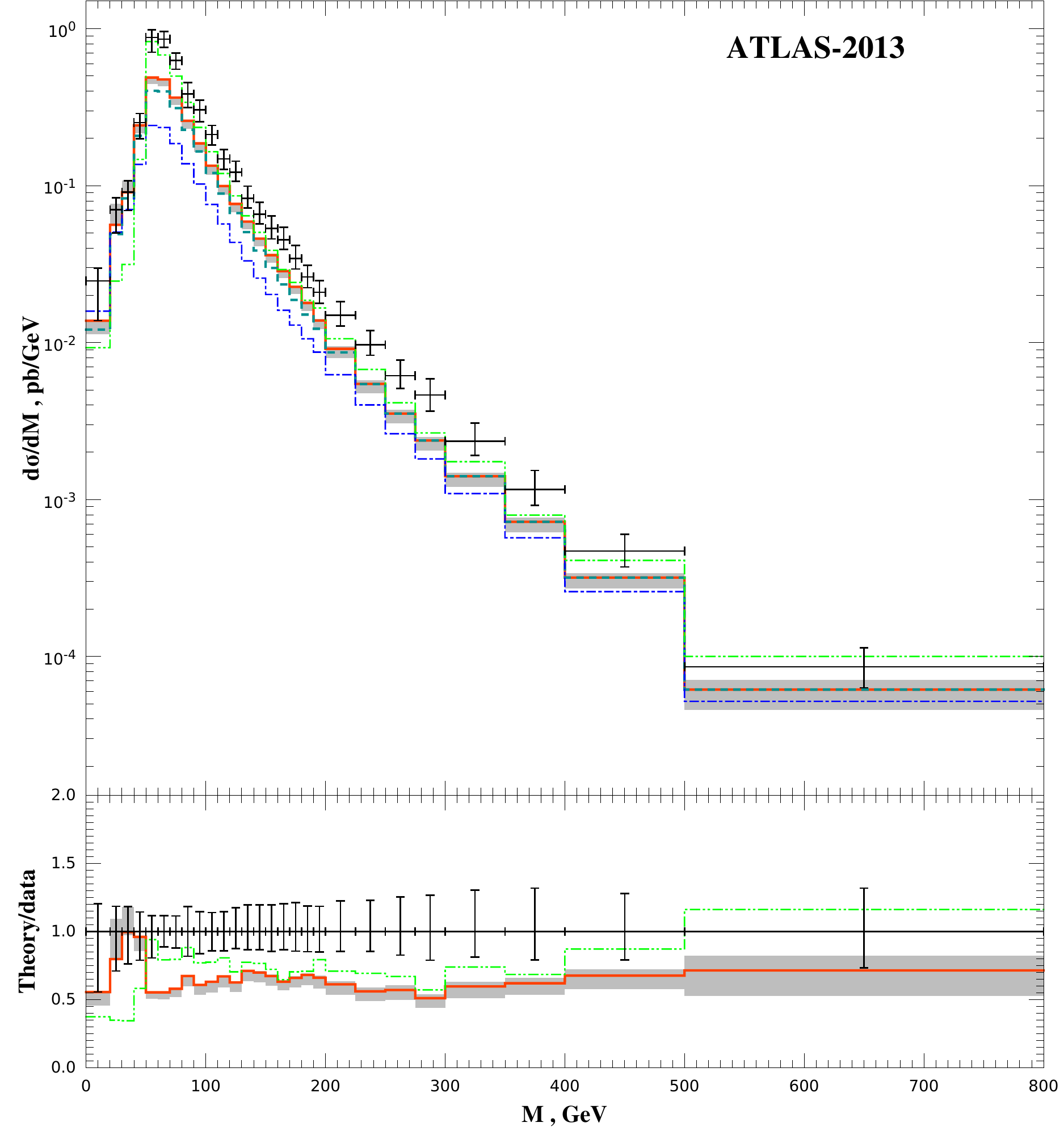}
  \caption{Diphoton invariant mass spectrum for the ATLAS-2013 dataset. The notation for the curves is the same as in the Fig.~\ref{fig:CDF_pT}}\label{fig:ATLAS_M}
  \end{figure}

  For the numerical computations we use the modified KMR unPDF~(\ref{eqn:KMR}) with the MSTW-2008 NLO PDFs~\cite{MSTW_2008} as the collinear input.
   Also, the value for the fine-structure constant $\alpha=1/137.036$ was used in the calculations together with the NLO formula for the $\alpha_s$
    with $\alpha_s(M_Z)=0.12018$ and flavor thresholds at $m_c=1.4$ GeV and $m_b=4.75$ GeV. The choice of the factorization and renormalization
    scales $\mu_R=\mu_F=\xi M$ commonly used in the literature~\cite{CDF_data_2012, ATLAS_data_2013, Diphox, 2gNNLO} was adopted,
    where the default value for $\xi=1$ and the values $\xi=2^{\pm 1}$ where used to estimate the scale uncertainty of the calculation, which is indicated in the Figs.~\ref{fig:CDF_pT} -- \ref{fig:ATLAS_M} as the gray band. The numerical computations where performed mostly using the \texttt{Suave} adaptive Monte-Carlo integration algorithm with the cross-checks against the results of \texttt{Vegas} and \texttt{Divonne} algorithms  implemented in the \texttt{CUBA} library~\cite{CUBA}.

\begin{figure}
  \includegraphics[width=0.45\textwidth]{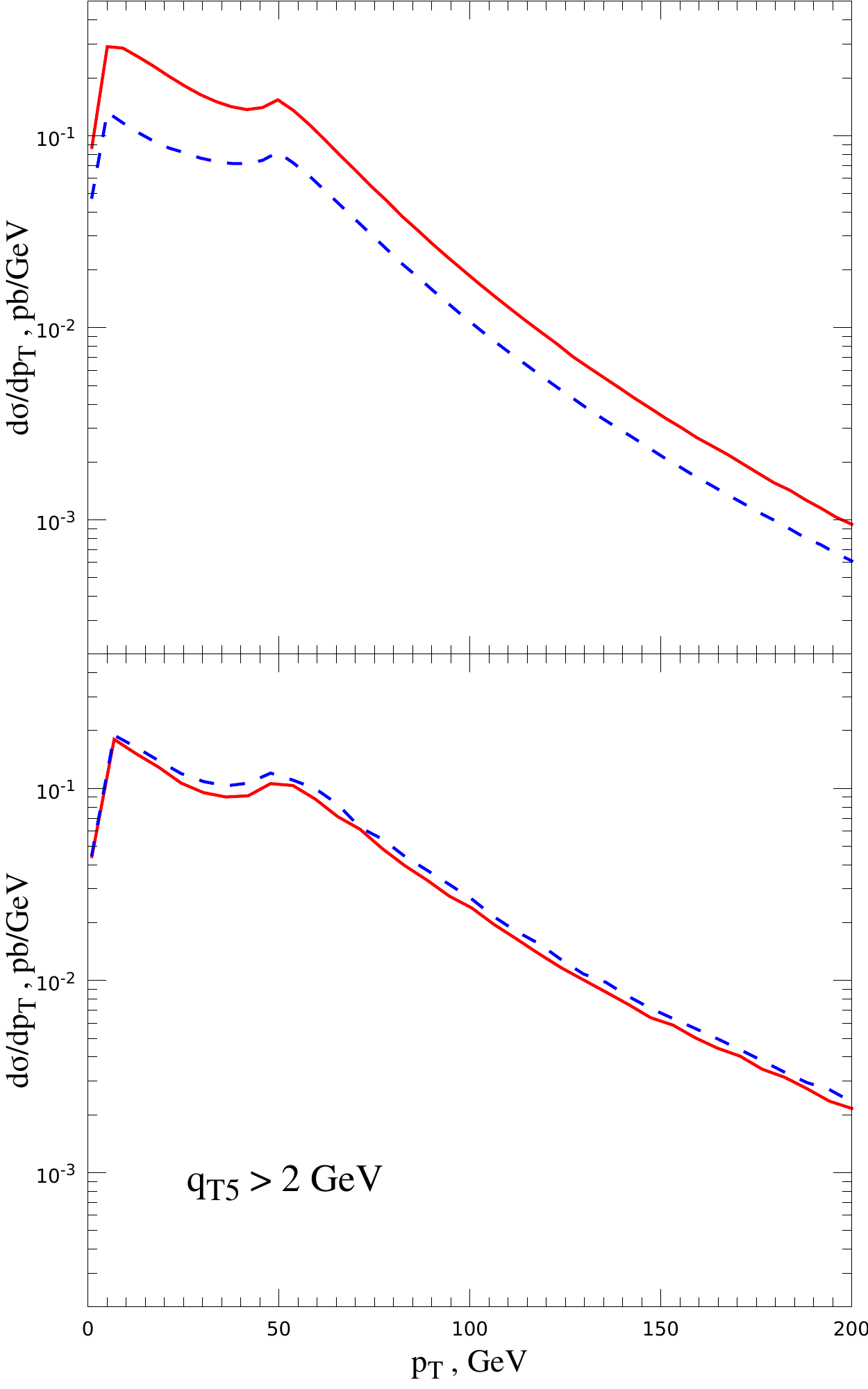}
  \includegraphics[width=0.45\textwidth]{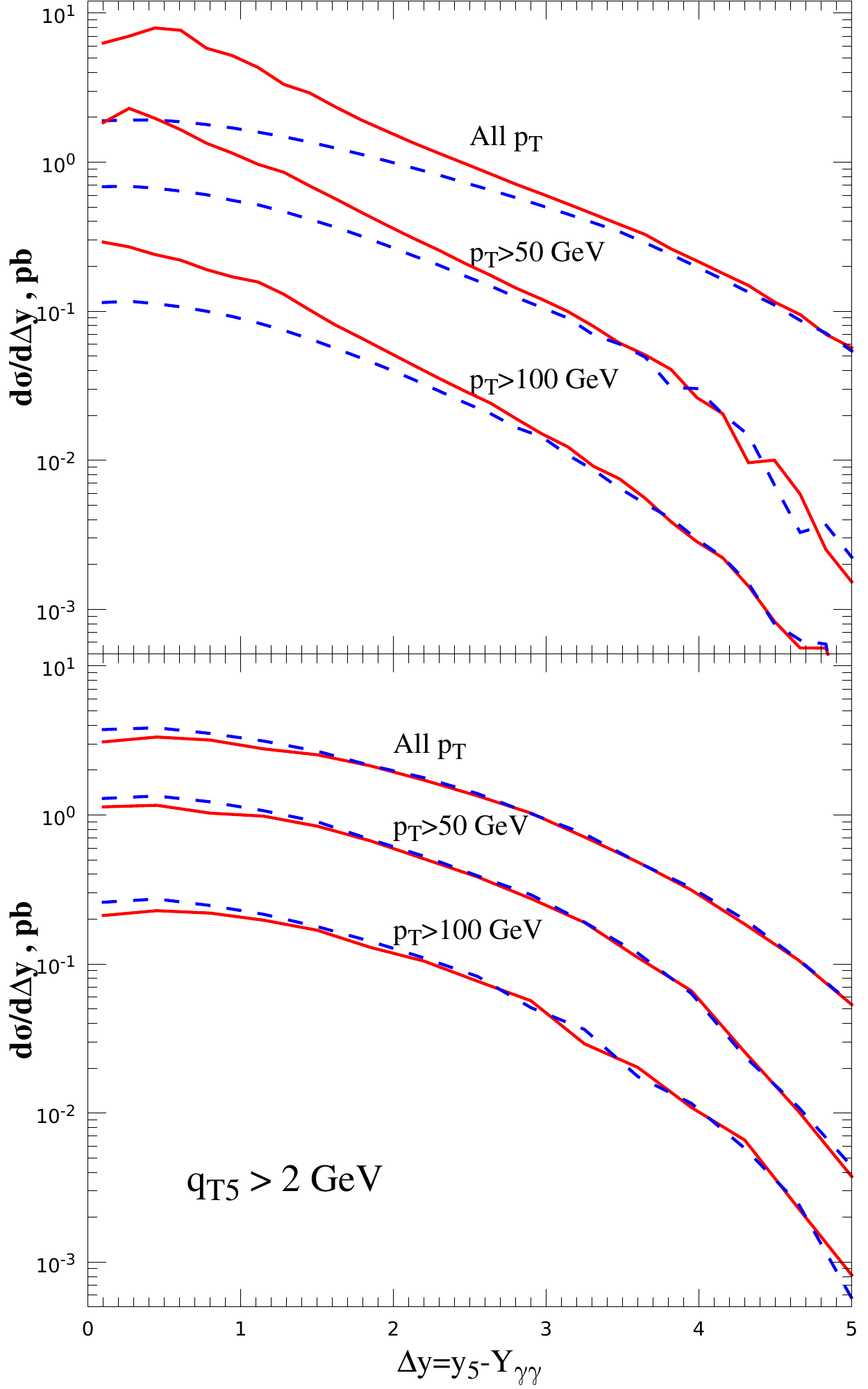}
  \caption{The comparison of the $p_T$ (left column) and $\Delta y$ (right column) spectra fpr the NLO QMRK contributions (solid lines) with the corresponding mMRK subtraction terms (dashed lines), upper panels -- subprocess (\ref{proc:QRgagaq}), lower panels -- subprocess (\ref{proc:QQgagag}).}\label{fig:NLO_subtr}
  \end{figure}

  Before the presentation of the comparison of our calculations with experimental data, let us discuss the contribution of
  $2\to 3$ subprocesses (\ref{proc:QRgagaq},\ref{proc:QQgagag}) after the subtraction of double counting, discussed in the Sec.~\ref{sec:realNLO}.
  In the left panel of the Fig.~\ref{fig:NLO_subtr} the $p_T$-spectra for the NLO contributions~(\ref{proc:QRgagaq}, \ref{proc:QQgagag}) in the ATLAS-2013 kinematical
  conditions~(see the second column of the table~\ref{tab:kinem}), is presented together with the corresponding mMRK subtraction contribution.
  For the CDF-2012 kinematics of the Ref.~\cite{CDF_data_2012} (first column of the table~\ref{tab:kinem}) the qualitative picture is the same.

\begin{table}
\begin{ruledtabular}
\begin{tabular}{cc}
$p\bar{p}$, CDF-2012~\cite{CDF_data_2012} & $pp$, ATLAS-2013~\cite{ATLAS_data_2013} \\
\hline
$\sqrt{S}=1960$ GeV & $\sqrt{S}=7000$ GeV \\
$q_{T3,4}\geq 15,\ 17$ GeV & $q_{T3,4}\geq 22,\ 25$ GeV \\
$|y_{3,4}|\leq 1.0$ & $|y_{3,4}|\leq 1.37$, $1.52\leq |y_{3,4}|\leq 2.37$ \\
$R=0.4$, $E_T^{(ISO)}=2$ GeV & $R=0.4$, $E_T^{(ISO)}=4$ GeV
\end{tabular}
\end{ruledtabular}
\caption{Kinematical conditions for the CDF and ATLAS datasets.}\label{tab:kinem}
\end{table}

  From the Fig.~\ref{fig:NLO_subtr} one can observe, that mMRK subtraction term reproduces the exact contribution of the
  NLO subprocess (\ref{proc:QQgagag}) with the $10\%$ accuracy and constitutes more than $50\%$ of the cross section of
  the subprocess (\ref{proc:QRgagaq}) for the $p_T>50$ GeV. As the right panel of the Fig.~\ref{fig:NLO_subtr} shows,
  for the subprocess~(\ref{proc:QRgagaq}) the significant deviation from the mMRK asymptotics starts only for
  $\Delta y=y_5-Y_{\gamma\gamma}< 2.0$ while for the larger values of $\Delta y$, the the QMRK $2\to 3$ cross
  section is well described by the mMRK asymptotics. For the subprocess (\ref{proc:QQgagag}) the $2\to 3$ cross section is
  reproduced by the mMRK approximation for all values of $\Delta y$. Consequently, more than $50\%$ of the cross section
  of the subprocess~(\ref{proc:QRgagaq}) and almost all contribution of the subprocess~(\ref{proc:QQgagag})
  will be canceled by the subtraction term. Having this in mind we do not include the contribution of the subprocess~(\ref{proc:QQgagag}) in the further calculations.

The squared amplitude for the subprocess (\ref{proc:QRgagaq}) can be
safely integrated from $q_{T5}=0$ in (\ref{cs:2to3}). The cross
section for the subprocess (\ref{proc:QQgagag}) is also finite, but
for the small values of $q_{T5}$, most of the cross section is
accumulated at $t_{1,2}\sim 1$ GeV$^2$, which is nothing else than the manifestation of the usual infrared singularity for the radiation of the soft gluon. For this reason the cutoff $q_{T5}>2$ GeV was imposed to produce the lower panel of the Fig.~\ref{fig:NLO_subtr}.
The dependence of the cross section on the small-$t$ behavior of the unPDF is unphysical and will be canceled away by the NLO real-virtual interference contribution.

  Now we are in a position to compare the predictions of our model with the experimental data of the refs.~\cite{CDF_data_2012, ATLAS_data_2013}.
  For the comparison, we choose three main observables, $d\sigma/dp_T$, $d\sigma/d\Delta\phi$, as ones, where the fixed-order calculations in the CPM are experiencing the greatest difficulties and $d\sigma/dM$ as the benchmark CPM observable, for which the multi-scale nature of the process under consideration is less important.  The comparison for the other observables will be discussed elsewhere.

  In the Fig.~\ref{fig:CDF_pT} the $p_T$-spectra of the photon pair, measured by the CDF Collaboration is presented. For this dataset,
  three data samples are provided, the inclusive one and two data samples with the additional kinematical constraint $p_T< M$ or $p_T> M$ imposed.
  One can note, that the inclusive data and data for the $p_T<M$ are well described for the $p_T>25$ GeV by the sum of the LO contribution (\ref{proc:QQgaga})
  and NLO contribution (\ref{proc:QRgagaq}) after the mMRK subtraction. For the $p_T>M$ data are well described by
  our prediction for all values of $p_T$, and the NLO contribution is, in fact, negligible. The contribution of the box subprocess (\ref{proc:RRgaga})
   is only about $15\%$ of the cross section predicted at small $p_T$, and decreases with $p_T$ very fast, contributing significantly only
   for the $p_T<30$ GeV.

  For the $p_T<25$ GeV one can observe the deficit of the predicted cross section which reaches up to a factor of $5$ at the $p_T=7$ GeV.
  The region of small-$p_T$ corresponds to the kinematics of CPM, where the radiation of soft gluons and virtual corrections are dominating.
  We expect, that computation of the NLO real-virtual interference correction in the PRA will significantly reduce this gap. One of the advantages
  of PRA is that at NLO this correction is finite and can be considered separately from the real NLO corrections, which are the subject
  of the present study.

  The good description of the data for the $p_T>M$ region supports the self-consistency of our approach, since the NLO correction in this
  region is almost canceled by the mMRK subtraction terms and the contribution of the real-virtual NLO correction is expected to be small here.

  For the reader's convinience, in the figs.~\ref{fig:CDF_pT} -- \ref{fig:ATLAS_M}, we also have plotted the corresponding NLO CPM predictions. Data for these plots correspond to the \texttt{Diphox} predictions~\cite{Diphox}, presented in the CDF~\cite{CDF_data_2012} and ATLAS~\cite{ATLAS_data_2013} experimental papers. The contribution of the $gg\to \gamma\gamma$ subprocess is also included into this predictions, via \texttt{GAMMA2MC} program~\cite{Bern_gg_gaga}. Comparing the NLO CPM and NLO$^\star$ PRA predictions in the figs.~\ref{fig:CDF_pT} -- \ref{fig:ATLAS_M} one can conclude, that, NLO$^\star$ approximation in PRA can not describe the data on the $d\sigma/dM$-distribution due to the absence of the loop correction, which contributes mostly in the back-to-back CPM-like kinematics. But for the configurations far away from the CPM kinematics, NLO$^\star$ PRA describes data substantially better than NLO CPM, especially at the LHC. Moreover, in this region, the NLO$^\star$ PRA prediction is dominated by the LO term, which demonstrates the better stability of PRA predictions for the kinematics far away from CPM one. Inclusion of full NLO corrections should also improve the agreement in the CPM region.    

  In the left panel of the Fig.~\ref{fig:ATLAS_pT} one can observe the same qualitative features as in the left panel of the Fig.~\ref{fig:CDF_pT},
  despite the fact, that we have moved from Tevatron to the LHC with it's $3.6$ times larger energy, and switched to $pp$ collisions instead of $p\bar{p}$ ones. The NLO subprocess~(\ref{proc:QRgagaq})
  is more important at the LHC than at the Tevatron, contributing significantly up to $p_T=200$ GeV.

  In the Fig.~\ref{fig:CDF_PHI} and right panel of the Fig.~\ref{fig:ATLAS_pT} the $\Delta\phi$-spectra for the Tevatron and LHC are presented.
  In the both figures one can observe a good agreement of our predictions with data for $\Delta\phi<1.5$ which corresponds
  to the high deviation from the back-to-back kinematics for the photons. In this region the NLO correction is manifestly
  subleading, as it was for the $p_T$-spectrum. The good description of the Tevatron data for the $p_T>M$ case is also there, as well,
  as the deficit of the predicted cross section for the back-to-back kinematics.

  As for the $M$-spectra of the Figs.~\ref{fig:CDF_M} and \ref{fig:ATLAS_M}, one certainly expects the deficit of the calculated
  cross section for the most values of $M$ due to the deficit of the cross section for the CPM kinematics, observed earlier, since most of the total cross section is accumulated near the CPM configurations. However, in the region of $M$ below the peak, the data are well described,
  demonstrating that PRA is suitable for the description of the effects of kinematical cuts. Once again, we observe the good description
  of the $M$-spectrum for the $p_T>M$ subset of the Tevatron data.

  The contribution of the quark-box subprocess~\ref{proc:RRgaga} to the $M$-spectra is found to be only about $8\%$ of the observed cross section in the peak,  and $18\%$ of the predicted cross section, both for the CDF-2012 and ATLAS-2013 kinematics. This result is $20-30\%$ smaller than the usual CPM estimate~\cite{2gNNLO}, which is in accordance with the findings of the Ref.~\cite{KNS_photon_jet} where it was shown, that the space-like virtuality of the initial-state partons suppresses
  the $\gamma R\to\gamma g$ contribution with respect to CPM expectation.

\section{Conclusions}
\label{sec:conclusions}

  In the present study, the pair hadroproduction of prompt photons is considered in the framework of PRA with tree-level NLO corrections
   (\ref{proc:QRgagaq}, \ref{proc:QQgagag}) and NNLO quark-box subprocess (\ref{proc:RRgaga}) taken into account. The procedure of localization
    in rapidity of the tree-level NLO corrections to avoid the double counting the real emissions between the hard-scattering part of the
     cross section and unPDF is proposed in the Sec.~\ref{sec:realNLO}. As a consequence of this procedure, the real NLO corrections are
      put under quantitative control, and their contribution was found to be numerically small at high $p_T$ or in the kinematical region $p_T>M$.
       The kinematical region $p_T>M$ is interesting for the further theoretical and experimental study, as an ideal testing site for the PRA,
        where the MRK between the ISR and the hard subprocess is dominating.
  The contribution of the quark-box subprocess~(\ref{proc:RRgaga}) was found to be about $8\%$ of the observed cross section in the peak of
  the $d\sigma/dM$ distribution, which is a bit smaller than the CPM estimate~\cite{2gNNLO} due to the space-like virtuality of the
  initial-state partons, similarly to the results of the  Ref.~\cite{KNS_photon_jet}.

\section*{Acknowledgements}
This work was supported by Russian Foundation for Basic Research
through the Grant No~14-02-00021.  The work of M.~A.~N. was also
supported by the Graduate Students Scholarship Program of the
Dynasty Foundation and by the Grant of President of Russian
Federation No~MK-4150.2014.2.
  Authors would like to thank M.~G.~Ryskin, G.~Watt and E.~de~Oliveira for providing to us their numerical codes for the calculation of the KMR unPDF
  of the Ref.~\cite{KMR_NLO}. M.~A.~N. would like to thank the Dept. of the Phenomenology of the Elementary Particles of the II Institute
  for Theoretical Physics of Hamburg University and personally Prof. B.~A.~Kniehl for their kind hospitality during the initial stage of
  this work and the computational resources provided.

\end{document}